# 6. THE INTERACTION OF THE ACTIVE NUCLEUS WITH THE HOST GALAXY INTERSTELLAR MEDIUM

*G. Fabbiano & M. Elvis*
*Center for Astrophysics | Harvard & Smithsonian*
*Cambridge, Massachusetts, USA*
*gfabbiano@cfa.harvard.edu*, *melvis@cfa.harvard.edu*

**ABSTRACT**

Imaging X-ray spectroscopy of nearby AGNs, mostly with Chandra, has shown that extended soft (<2.5 keV) emission-line dominated X-ray biconical structures, of kiloparsec scale, are widespread in highly absorbed Compton Thick (CT) AGNs. The X-ray emission is complex, requiring both photoionized and shock-ionized gas. It originates from high ionization regions and is surrounded by cocoons of low ionization narrow line emission regions (LINERS). Biconical ~3-6 keV continuum and 6.4 keV Fe K$\alpha$ emission has been detected, contrary to the standard AGN model expectation that would confine this hard emission to the pc-size nuclear absorbing torus. Extended emission in the cross-cone direction also requires modifications to the AGN standard model. A porous torus, with a significant fraction of escaping AGN continuum, and/or jet interaction with ISM creating a "blow-back" towards the nuclear region seem to be required. Here we discuss these results and their implications for both the AGN model and our understanding of AGN feedback. The finding of hot and highly photoionized gas on 10s parsecs to several kiloparsec scales demonstrates that all three feedback mechanisms are at work: radiation affects the inner molecular clouds of the host on a ~1 kpc scale; shocks of relativistic jets with the host ISM a few kpc from the central AGN; and photoionization of the ISM via winds on scales from pc to multiple kpc. These results demonstrate that X-rays are needed to develop a complete picture of AGN/host interaction along with radio continuum, mm and sub-mm molecular line emission, and optical/near-IR emission lines.

## 6.1 INTRODUCTION AND CHAPTER OUTLINE

Active Galactic Nuclei (AGNs) are normally considered the quintessential 'point' sources. Their electromagnetic emission is believed to be the result of the gravitational accretion of circumnuclear material (stars and gas) onto central supermassive black holes at the nuclei of galaxies (Rees 1984). Yet they also show extended structures from parsec to several kiloparsec scales in the radio, optical, and X-ray bands. This chapter explores what is known about the extended X-ray emission of AGNs and what these features imply for the structure of the AGN and its influence on its surrounding host galaxy, known as AGN feedback.

The first AGNs recognized as a new class of objects are the *quasars*, so named for their stellar, point-like, appearance (Schmidt 1963). Their rapid X-ray variability by factors of two or more on timescales from years (for luminous quasars, Zamorani et al. 1984, Giveon et al., 1999) down to days or minutes (for low luminosity AGNs, Elvis 1975, Boller et al., 1993, Papadakis & Lawrence 1995) points to light-crossing times that are similarly short. The relativistically beamed *blazars*, a family of variable, polarized AGNs (Angel & Stockman 1980), for which this argument was first made, have their variability timescales shortened by an order of magnitude (by the relativistic factor Γ, where Γ ~ 10, Angel & Stockman 1980). Instead, radio-quiet quasars show a linear relation between their X-ray luminosity and their optical emission line luminosity (Elvis et al., 1978, Heckman et al. 1985). The emission lines are not relativistically Doppler-shifted, so neither are the X-rays and therefore the light-crossing time argument can be used to estimate the size of the emitting region. These estimates, plus arguments based on both X-ray eclipses (Risaliti et al. 2005) and on General Relativistically broadened Fe-K lines (Fabian, et al., 2014), all restrict the bulk of AGN X-ray emission, at least in moderate luminosity AGNs, to originate from a region with dimensions of just a few Schwartzschild radii[1].

Yet AGNs are not X-ray point sources. There is a non-varying component to the soft X-ray emission (Weaver et al., 1994), which turns out to be extended on large scales and has been directly imaged in ever more detail, most notably with the Chandra X-ray Observatory (Weisskopf et al. 2002). This chapter describes the state of our knowledge of this extended X-ray emission in AGNs.

This extended emission is mostly collimated (and referred to as a 'bicone') and is a marker of both illumination and outflows from the central AGN, and of interactions with the host galaxy interstellar medium (ISM). From this emission we can probe the mechanisms leading to X-ray emission and estimate the strength of the AGN feedback onto star formation in the host galaxy.

We only briefly deal with 'radio-loud' AGNs in this chapter. Radio-loud AGNs are typically hosted in early type (elliptical and lenticular) galaxies and are rarer, and so typically more distant, than radio-quiet AGNs, meaning that only larger-scale structures can be resolved. They are discussed more fully in Chapter 3 from the point of view of feedback.

The outline of this Chapter is as follows:

- Section 6.2 introduces the theoretical and observational landscape that motivates the Chandra high resolution spatial / spectral imaging of AGNs, including: AGN feedback on galaxy evolution (6.2.1), multi-wavelength observations (6.2.2), and the AGN Standard Model (6.2.3).
- Section 6.3 summarizes the history of the extended X-ray emission in AGNs and reviews the X-ray spectral components found in CT AGNs (6.3.1).
- Section 6.4 presents the Chandra results on the extended soft (<~2.5 keV) bicone emission components of AGNs, starting with a compilation of results so far (6.4.1), followed by a discussion of the imaging techniques used to highlight these components (6.4.2), a summary of the broad-band (0.3-2.5 keV) X-ray morphology (6.4.3), of the morphology of narrow-band imaging centered on X-ray emission lines (6.4.4) and of the spectral features of this emission (6.4.5). We conclude with a discussion of the spatial morphology of emission lines in these features (6.4.6).

---

[1] Where the Schwartzschild radius, $R_g = 2GM/c^2$, = $3 \times 10^{13}$ cm (2 AU) for M = $10^8$, where M is the mass of the black hole in solar masses, and G and c are the gravitational constant and the speed of light respectively.

– Section 6.5 discusses the Chandra discovery of extended hard continuum and neutral Fe Kα bicone emission, as well as localized Fe XXV emission (6.5.1).
– Section 6.6 presents evidence of extended X-ray emission in the cross-cone direction and discusses its implications.
– Section 6.7 reports high resolution studies of the innermost AGN surrounding that can be imaged with Chandra and of their relevance for understanding the obscuring "parsec-scale" torus and the physical properties of these regions.
– Section 6.8 discusses the implications of Chandra AGN observations for understanding the time-evolution of these objects.
– Section 6.9 revisits the implications of these results for AGN feedback.
– Section 6.10 summarizes the main results discussed in this Chapter and looks at future developments in this field.

## 6.2 THEORETICAL AND MULTI-WAVELENGTH OBSERVATIONAL BACKGROUND

Before discussing the observations and properties of the extended X-ray emission of AGNs, we summarize in this section the scientific background relevant for understanding these new observations. We start with a short review of the present understanding of galaxy evolution and the importance of AGN feedback in this process (Section 6.2.1), then discuss the multiwavelength observational evidence of the interaction of the AGN with the interstellar medium of the host galaxy (Section 6.2.2), and conclude with a summary presentation of the 'Standard Model of AGN' resulting from both multi-wavelength and early X-ray observations (Section 6.2.3).

### 6.2.1 Galaxy Evolution and Feedback

Galaxy formation and evolution models have long needed a source of energy input to explain the galaxy luminosity function. Large scale structure simulations in the ΛCDM cosmology (e.g., Benson et al., 2003), using only gravity, produce a power-law mass function for galaxies, while the observed mass function follows a Schechter law: a broken power-law (Schechter 1976) that matches the ΛCDM predictions only at the inflection point, with deficits at both high and low luminosities (e.g., Kereš et al., 2009). At low luminosities the observed deficit of galaxies can be readily explained by supernovae pushing on the ISM and accelerating it to above escape velocity; this is called supernova feedback and is closely linked to star formation to provide sufficient numbers of OB stars (Kereš et al., 2009). In contrast, the deficit at high luminosities requires more energy than supernovae alone can provide. Tapping the energy from central AGNs is appealing as they release more than enough to unbind the ISM of the host galaxy: the bulge binding energy, $E_{bulge} \sim M_{bulge}\,\sigma^2$, where σ is the velocity dispersion ~400 km $s^{-1}$, while growing the black hole of mass $M_{bh}$ radiates $E_{bh} \sim \eta M_{bh} c^2$, where η is the radiative efficiency ~0.1. Since $M_{bh}$ ~1.5 x $10^{-3}$ $M_{bulge}$, $E_{bh}/E_{bulge} > 80$ (Fabian 2012). Even a small fraction of the radiative power coupled to the host ISM can have a profound effect on that medium; this is called AGN feedback.

An additional impetus to look for feedback between the AGN and the host galaxy is the strong observed correlation between the mass of the bulge of the host galaxy and that of the central black hole, the $M_{BH}$-σ relation (e.g., Gultekin et al., 2009). The correlation seems to imply that growth of the bulge mass and the growth of the black hole are connected, with $M_{bh} \sim 1.5 \times 10^{-3} M_{bulge}$ (Merritt & Ferrarese 2001).

There are three possible avenues for AGN feedback onto the host ISM: radiation, jets, and winds (Fabian 2012). There is clearly plenty of radiative power available, but most of the radiation escapes. The relativistic, highly collimated, jet power is adequate for radio loud AGNs, especially for the central dominant galaxies in clusters and groups of galaxies (see Chapter 3 by Nardini, Kim and Pellegrini.) In X-ray binaries and tidal disruption events, jets form only at super-Eddington ratios (Curd & Narayan 2019, Blandford, Meier and Readhead 2019). Jets may likewise be relatively short-lived in AGNs. Indeed, ~90% of AGNs are radio quiet (Kellerman et al., 1989; Macfarlane et al., 2021), and accrete well below the Eddington limit (Kollmeier et al., 2006, Steinhardt & Elvis, 2011).

Winds are sub-relativistic, less highly collimated outflows. Unlike radio jets they are common, and potentially universal, in radio quiet AGNs. Whether AGN winds carry sufficient power to create AGN feedback has been unclear. The question is how to couple the AGN power to the ISM efficiently enough to stop star formation in the host. Early calculations simply asked how much energy was needed to unbind the ISM. This takes about 5% of the AGN power (Di Matteo et al., 2005). Another approach is to disrupt the molecular clouds from which new stars form. An AGN wind can do this using only ~0.5% of the AGN power (Hopkins and Elvis, 2010). More realistic simulations at higher resolution and including more physics may change this number (e.g., FIRE-2, Su et al., 2021).

A significant impediment to studying AGN feedback onto host galaxies is the mismatch in their timescales (Hickox et al. 2014). AGN lifetimes are $\sim 10^7$ yr (Martini 2004) with factors of several variability on much shorter timescales, down to a year or even less, while star formation operates on a $\sim 10^8$ yr timescale. This mismatch will weaken any correlations between the two variables in samples of galaxies or AGN.

An alternative way to study feedback is via imaging studies of particular cases. These studies can evaluate the importance of AGN feedback on the host ISM as it is happening. X-ray imaging spectroscopy is a particularly useful technique to study all three feedback mechanisms. Characteristic wind outflow speeds are ~1000 km $s^{-1}$, which will thermalize in shocks to give gas temperatures of $\sim 10^7$ K. Plasmas at these temperatures emit X-rays in the ~1 keV range, making soft X-ray imaging particularly valuable. Jet/ISM interactions should lead to shocks of at least the same temperature. Irradiation of molecular clouds by the AGN would produce fluorescence signatures in the hard X-ray band, resulting in the 6.4 keV neutral Fe K$\alpha$ line (e.g., Ueno et al., 1994).

### 6.2.2 Multi-wavelength Imaging of Radio-quiet AGN Interactions with Host Galaxies

Extended emission due to the presence of an AGN has been found in several spectral bands – radio, millimeter and sub-millimeter, infrared, optical, and X-ray. Blueshifted absorption lines unambiguously indicate the presence of outflows in AGNs. These absorption lines in the UV and X-ray bands are summarized briefly by Fabian (2012) and Fiore et al. (2017). We will discuss instead the X-ray *emission* evidence in the rest of this chapter. Here we summarize the evidence from emission lines in the other spectral bands.

Almost no AGN is truly radio silent; radio emission is present even in radio quiet AGNs (Kellerman et al., 1989, Macfarlane et al., 2021). Morphologically these radio structures look like ~kpc-scale miniatures of the powerful radio loud AGNs (Wilson & Willis, 1980) They have typical luminosities of $10^{20}$ – $10^{23}$ W $Hz^{-1}$ (Ulvestad & Wilson 1989), ~$10^{-5}$ of the optical/UV luminosities of these objects (Elvis et al., 1994). The direction of these radio structures is not correlated with their host galaxy disk orientation (Kinney et al., 2000).

Optical extended regions of [OIII] emission were found early on in AGN studies (Walker 1968), but came into their own with the higher angular resolution capabilities of the Hubble Space Telescope (HST). Their large [OIII]/Hα ratios distinguish these regions from star formation regions and require the high energy ionizing radiation of an AGN. These regions are common in AGNs (Schmitt et al., 2007). They often, but by no means always, have a clear bi-conical shape (Tadhunter and Tsvetanov, 1989, Fischer et al., 2013.) Such a shape was first seen as an illumination pattern in which the inner obscuring torus of the Unified Scheme collimates the AGN light and so they were called "ionization cones" (Wilson 1996). Later long-slit spectroscopic observations with HST/STIS suggest that these structures are coherent matter bounded, hollow, outflows (Crenshaw and Kraemer, 2000, Fischer et al., 2013) and so are, noncommittally, now usually called bicones. Modeling of the kinematics of these bicones gives kinetic powers as a fraction of bolometric luminosity >0.3% and some lie in the 0.5% - 5% range. Moreover, quite sudden (<100 pc-scale) deceleration at distances of 0.1 – 1 kpc from the nucleus is seen in some of these outflows implying a loss of kinetic energy, presumably to interaction with the host ISM (Fischer et al., 2013). Where this kinetic energy goes is unclear from optical observations.

Given the high temperatures associated with AGN outflows it was initially surprising that cold molecular outflows in CO were found at comparable speeds in numerous AGNs (e.g., MRK 231, Feruglio et al., 2010, Fiore et al., 2017). In MRK 231 the mass outflow rate, presumably of entrained ISM, is ~700 $M_{sol}$/year and the kinetic power is ~5% of the AGN luminosity. Fiore et al. (2017) find values of ~1% - ~10% for a sample of ~10 molecular winds. How much of this power is coupled to the ISM? The molecular gas could well be ISM gas entrained in the winds. Cool material within the outflow direction is also found as thermal hot dust emission in some AGNs (Asmus et al., 2016), although dust may form in the outflow itself (Elvis, Marengo, & Karovska, 2002). The interaction of the cold molecular ISM with the AGN is not limited to outflows. Nuclear inflows are also suggested in some cases (Storchi-Bergmann et al., 2007, Feruglio et al 2020). These inflows may be a path by which the galaxy powers the nuclear black hole, completing the feedback loop by controlling the black hole accretion rate.

Finally, in our Milky Way galaxy, the Fermi bubbles seen in GeV gamma-rays (Su, Slayter & Finkbeiner 2010) and the X-ray bubbles seen by eROSITA (Predehl et al. 2020) may be a similar "bicone" structure, due to an active episode in SgrA* a few Myr ago (Dobler et al., 2010, Nicastro et al., 2016). Also, smaller outbursts from SgrA* are seen in X-ray reflection and fluorescence off molecular clouds close to SgrA* (Koyama et al. 1996; Ponti et al. 2015; Churazov et al. 2017a, 2017b). Similar features should be found in the more luminous AGNs.

### 6.2.3 The 'Unified Scheme' of AGNs

Early AGN optical spectroscopy recognized two major classes (Khachikian and Weedman, 1974): type 1 AGNs, with broad emission line spectra (FWHM ~ 0.02 c) and are typically more luminous, and type 2 with narrow emission line spectra (FWHM ~ 0.002 c) and are mostly less luminous. The Unified Scheme posited that type 2 (narrow emission line) AGNs were heavily obscured type 1 (broad emission line) AGNs seen through an optically thick torus. The inner radius of the obscuring torus scales as $L_{uv}^{1/2}$ and is set by the sublimation temperature of the dust grains, ~1500 K for graphite. For a UV luminosity of $10^{46}$ erg $s^{-1}$ (similar to that of 3C 273) this comes out to be ~1 pc (e.g., Barvainis, 1987). This size is larger than the broad line emitting region (BLR), but smaller than the narrow line emitting region (NLR). The obscuration by the dust and gas in the torus could explain why type 2 AGNs were so much fainter in X-rays than type 1 AGNs and lacked broad emission lines (Lawrence and Elvis 1982) and why broad emission lines could be seen in polarized light in some type 2 AGNs (Antonucci and Miller, 1985). The torus configuration for the obscuring material allowed an AGN to be seen as a type 1 from angles near the torus axis, and a type 2 from angles near the plane of the torus. Hence in type 2 objects any radiation or outflow would be collimated into bi-conical regions, as observed (Section 6.2.2).

A major motivation for the Unified Scheme was the early discovery that some AGNs were highly obscured in X-rays by gas and dust with equivalent hydrogen column densities from ~$10^{22}$ $cm^{-2}$ up to a few x $10^{23}$ $cm^{-2}$ (Ives, Sanford and Penston, 1976, Mushotzky et al., 1980). As the photoelectric cut-off energy increases with $N_H$, the upper bound to the amount of obscuration was limited by the ~2 - 10 keV energy range of the sensitive X-ray missions to between a few x $10^{21}$ and a few x $10^{23}$ $cm^{-2}$, while the $\tau_{es} = 1$ Compton scattering column density is 3 x $10^{24}$ $cm^{-2}$. As AGN identifications grew in number it was realized that obscuration is common and is correlated with optical reddening (Ward et al., 1978, Mushotzky 1982). The degree of obscuration in both optical and X-rays was found to depend on the AGN luminosity (Lawrence and Elvis 1982, Netzer 2015). The torus soon became an accepted component of AGN structure (Urry and Padovani 1995), despite a number of open questions (Krolik and Begelman, 1988).

The torus could well be Compton Thick ($\tau_{es} > 1$). Yet soft X-rays were seen from type 2 AGNs, leading to the idea that perhaps the torus had gaps in it, a "partial covering" model (Reichert et al. 1985). Alternatively, the big gap towards the torus poles must contain electrons extending above the torus in order to scatter the broad emission line photons and create the distinctive polarized optical spectrum (Antonucci and Miller, 1985). This region would scatter X-rays just as strongly, as Compton scattering is energy independent up to ~30 keV. As the torus has ~parsec extent, this scattered component would not be variable on short (<1 yr) timescales. However, X-ray imaging observations from the Einstein and ROSAT X-ray telescopes (see 6.3) made clear that there was, in addition, a much larger-scale (~kpc) extended X-ray component.

Thanks to X-ray missions sensitive to higher energies, notably the BAT instrument on Swift (15 - 150 keV), we now know of many Compton Thick, Type 2, AGNs (e.g., Marchesi et al., 2018). The first Compton Thick AGNs (CT AGNs) were found with the JAXA ASCA satellite from their strong (EW >~1 keV) neutral Fe-K emission lines at 6.4 keV (e.g., NGC1068, Ueno et al., 1994). The 6.4 keV line arises from fluorescence off of high column density gas. By chance the cross-section for fluorescence is comparable to that for Compton scattering (e.g., Tomblin 1972). In the Unified Scheme for AGN the obvious location for Compton thick material is the torus.

## 6.3 EARLY X-RAY OBSERVATIONS OF EXTENDED AGN EMISSION THROUGH CHANDRA

Despite the expectation that AGNs would be highly compact X-ray sources, the first arcsecond-class imaging X-ray telescope, the Einstein Observatory (Giacconi et al., 1979) demonstrated that much of the soft X-ray flux of the prototype Seyfert galaxy NGC 4151 was extended on kiloparsec scales, and was roughly aligned with both the weak double-lobed radio source and the extended narrow emission line axes (Elvis, Briel and Henry, 1983).

Hints that extended soft X-ray emission was common in nearby AGNs were found in further observations with Einstein (NGC 1566, NGC 2992, Elvis et al. 1990) and with the later ROSAT mission (NGC 1068, Wilson et al. 1992; NGC 2110, Weaver et al. 1995; possibly NGC 7582, Schachter et al. 1998.) Both Einstein and ROSAT (Trümper 1982) were limited by their ~5 arcsecond resolution and by their lack of useful energy resolution in their microchannel plate imagers. This made the identification of the emission processes for these extended features uncertain. Their high surface brightness did rule out a normal galaxy hot ISM or unresolved point sources, pointing to an AGN related origin. Their alignment with the emission line regions, now recognized as bicones, and the weak radio jets also pointed to an AGN origin, likely related to outflows (see Section 6.2.2). But distinguishing between thermal, non-thermal, reflection/fluorescence, and photoionization origins was not possible. This left the kinetic power in the outflows undetermined.

With the launch of CCD imagers with good ($E/\Delta E \sim 10$) energy resolution at the focus of the large, higher angular resolution, mirrors on Chandra (Weisskopf et al. 2002) and XMM-Newton (Jansen et al., 2001) in 1999, extended soft X-ray components were clearly detected in nearby AGNs (Bianchi, Guainazzi and Chiaberge, 2006, Levenson et al. 2006). AGNs showing X-ray extent are typically highly obscured Compton Thick (CT) and/or narrow emission line-dominated in their optical spectra.

There are two distinct reasons for this special role of CT AGNs in the study of extended X-ray emission: (1) in the Unified Scheme (Section 6.2.3) we view CT AGNs side on, through the torus, so that the axis of the outflow bicones lies close to the plane of the sky, maximizing their projected extent; (2) the obscuring material cuts down the intensity of the central, unresolved point source of the AGN, acting as a natural coronagraph and allowing faint extended emission to be seen in the wings of the point spread function (PSF). In the case of Chandra this second reason dominates the selection of target AGNs, because the CCD readout time of the main Chandra instrument (ACIS) is too slow to count individually all the photons from high-flux sources.

### 6.3.1 The Spectral Components of CT AGN Emission

The observed X-ray spectrum of CT AGNs shows 3 or 4 distinct components:

1. Soft (0.3 - ~2.5 keV). This component is dominated by line emission; lines commonly seen in CCD spectra of CT AGNs include C V, O VII, O VIII, Ne IX, Ne X, Si XIII, Fe XVII-XIX (Section 6.4.4, Table 6-2). Grating spectra for a few CT AGNs display a similar variety of X-ray emission lines (Kaspi et al., 2001, Dadina et al., 2010). Soft band emission is largely extended (see above and Section 6.4.1); in the standard model, this soft component is due to nuclear photons escaping in the bicone direction and interacting with the host galaxy ISM.

2. Hard continuum (~3 – 6 keV); This energy range is dominated by a hard, usually variable amplitude, power-law continuum with a photon index, Γ ~ 1.7 (Mushotzky, Done, and Pounds, 1993). However, with Chandra the hard band continuum has been discovered to be sometimes extended (Section 6.5).

3. Fe-Kα lines (6.4 – 6.7 keV); In CT AGNs the most prominent is the neutral 6.4 keV Fe-Kα line, which in the standard model would be originating from fluorescence induced by the interaction of hard nuclear photons with the AGN torus. In this scenario, the 6.4 keV line should appear point-like in imaging space, but extended components have been discovered with Chandra (Section 6.5). The 6.7 keV Fe XXV line has also been detected in some AGNs and tends to be associated with thermal emission from shock-heated ISM (Section 6.5). Figure 1 illustrates the main spectral components in the ACIS spectrum of ESO 428-G014 from Chandra.

4. Compton hump (7 - ~40 keV); above the Fe-K edges at ~7 keV there are no other abundant elements to produce photoelectric absorption. This absence of absorption leads CT AGN X-ray spectra to recover their underlying power law in the ~7 keV - ~10 keV band, but in addition they typically show an excess over and above this power law. This excess is due to Compton scattering off an optically thick absorber, normally taken to be the torus. By ~40 keV electron recoil reduces this scattering according to the Klein-Nishina cross-section, so the spectrum drops back down to the power law (Awaki, H., et al., 1991). The Compton scattering excess is called the “Compton hump”. The amplitude of the Compton hump depends on the covering factor and geometry of the scattering material (Baloković et al., 2018). Most analyses use data with a large (~arcminute) beam size and assume that all the hard emission originates from an unresolved scattering region. Any extended hard emission, including iron fluorescence and Compton scattering, will affect the parameters of fits to the torus properties. Less obscured AGNs recover to the power law at lower energies as the iron, and other heavy element, column density is insufficient to produce photoelectric absorption at these energies.

5. The power law continues from ~40 keV to higher energies, ~140 keV - ~500 keV, where it eventually drops off exponentially (Baloković et al., 2020).

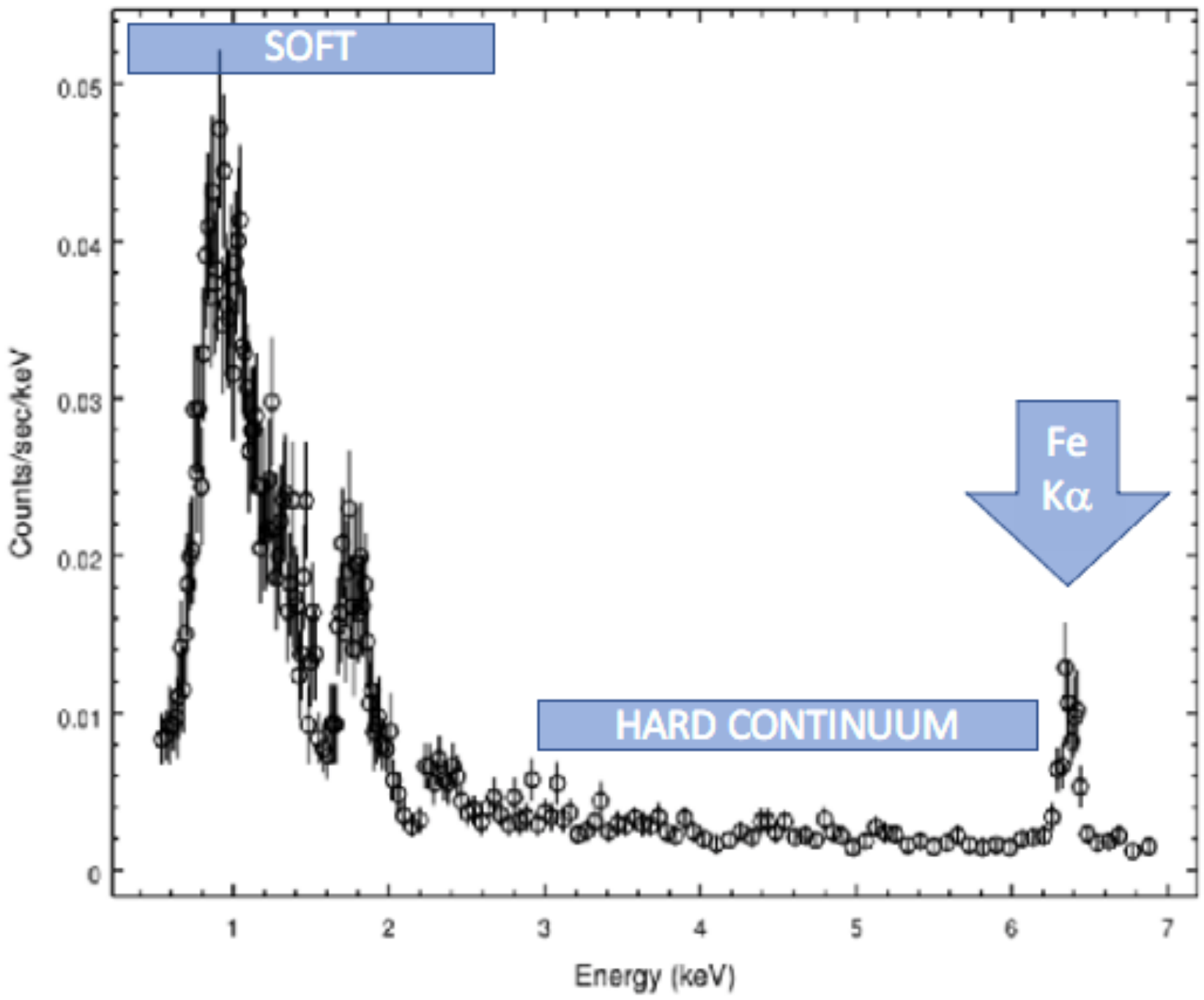

Figure 6-1: ESO 428-G014 ACIS spectrum showing the different components commonly seen in CT AGN X-ray spectra: the soft emission line dominated component (energy < ~2.5 keV), the hard continuum (~3.0 – 6 keV), and the prominent 6.4 keV Fe Kα neutral line at 6.4 keV.

## 6.4 CHANDRA IMAGING: THE SOFT COMPONENT

### 6.4.1 Prevalence of Extended X-rays

Table 6-1 lists, in order of distance from Earth, the 27 AGNs with well-studied soft (< 2.0 keV) X-ray extent in high S/N, high resolution, Chandra ACIS-S images at the time of writing. Note that exposure times can reach one or more days (1 day = 86.4 ks). Structures are seen that range over a factor ~100 in their physical size, from ~30 pc to ~3 kpc. The minimum resolvable size (~1/4”) goes from less than 5 pc to ~200 pc, depending on distance from us, and the outer limit to detectable extent is limited by surface brightness. Structure is seen on all accessible scales and so the true physical upper and lower limits remain unknown.

The luminosity of the extended soft X-ray structures spans a factor of nearly 1000 [log $L_{soft}$ (energy < 2 keV)= 38 – 40.9]. Imaging these structures with good signal-to-noise typically requires exposure times of at least 100 ksec with the CCD ACIS detector on Chandra.

The sample selection for these AGNs was primarily that they should be CT AGNs ($N_H > 10^{23.5}$ $cm^{-2}$) with the highest [OIII] luminosities in the Revised Shapely-Ames catalog (Risaliti, Maiolino & Salvati 1999), as $L_X$ correlates with L([OIII]). Long observations were proposed individually for those that had an indication of X-ray extent from short observations and/or had bright [OIII] extended narrow-line regions given the known correlation of [OIII] with soft X-ray luminosity (Heckman et al., 2005).

This selection process makes the prevalence of soft X-ray extent hard to judge, even among CT AGNs. The small surveys by Ma et al (2020) and Jones et al., (2021) filled in many of the gaps in the Chandra coverage of the Risaliti et al. (1999) sample. As such they provide a better estimate of soft X-ray extent prevalence. They indicate that as much as ~80% of CT AGNs have detectable soft X-ray extent.

Table 6-1: AGN with well-studied soft X-ray extent in order of distance, D.

| **AGN Host Type**[a] | **D Mpc** | **1” pc** | **ACIS exp., ksec** | **Nuclear log $N_H$ cm$^{-2}$** | **Radial extent**[b] **“, pc** | **Extended log($L_X$)**[c] **(erg s$^{-1}$) 0.5-2 keV** | **Hard or FeK extent?** | **References**[d] |
|---|---|---|---|---|---|---|---|---|
| NGC 4945 SBcd | 3.7 | 18 | 49 | 24.6 | 30, 540 | 38.0 | torus | Marinucci+ 2012 Schurch+ 2002 |
| Circinus SAb | 4 | 19 | 60 | 25.0 | 20, 380 | 38.8 | yes | Smith+2001 Arévalo+ 2014 Sambruna+ 2001 |
| NGC4258 SABbc | 7.2 | 34 | 35 | 23.0 | 120, 4100 | 40.3 | | Yang, Y.+2007 |
| NGC1386 SB0 | 12 | 57 | 100 | 24.8 | ~30, 1700 | 39.6 | yes | Jones+2021 |
| NGC 4151 SABab | 13.3 | 65 | 1.4 +180 HRC | 23.0 | 20, 1300 | 40.2 | | Wang+2011 Mushotzky+ 1980 Ives+1976 |
| NGC 1068 SAb | 14.4 | 72 | 40 | >24.5 | 6, 430 | 41.8 | yes | Wang+2012 Bauer+2015 Levenson+ 2006 |
| NGC 4388 SAb | 16.7 | 79 | 20 | 23.5 | 200, 16000 | 40.0 | yes | Iwasawa+ 2003 Yi+2021 |
| NGC 1365 SBb | 19 | 90 | 100 | >=23.3 | ~60, 5400 | 40.4 | | Wang+2009 Risaliti+2007 |
| NGC 5643 SABc | 21 | 100 | 114 | 24.2 | ~20, 2000 | 40.2 | yes | Jones+2021 |
| ESO 428-G014 SAB0 | 23.3 | 112 | 155 | >24.5 | 10, 1100 | 39.3 | yes | Fabbiano+ 2018 |
| NGC 5252 S0 | 33 | 157 | 60 | 22.3 | 15, 2350 | 40.1 | | Dadina+ 2010 |
| NGC 2110 SAB0 | 35 | 170 | 46 | 22.5 | 30, 5100 | 39.5 | | Fabbiano+ 2019 Evans+2006 |
| NGC 5347 SBab | 38 | 180 | 37 | | ~3, ~540 | 39.2 | No | Bianchi+2006 |
| ESO 137-G034 SAB0/a | 39 | 185 | 45 | 24.4 | ~6, 1100 | 40.0 | yes | Ma, J.+2020 |
| NGC 4500 SBa | 45 | 213 | 18 | 23.9 | 4.5, 960 | 39.7 | marginal | Ma, J.+2020 |
| NGC 1125 SB0/a | 47 | 224 | 53 | 24.2 | 5, 1100 | 39.7 | marginal | Ma, J.+2020 |
| NGC 3281 SAab | 48 | 233 | 9 | | 8, 1750 | 39.9 | yes | Ma, J.+2020 |
| IC 5063 SA0 | 51 | 240 | 270 | 23.5 | 15, 3600 | 40.0 | yes | Travascio+ 2021 |
| NGC 424 SB0/a | 51 | 241 | 15 | 24.2 | 8, 1930 | 40.3 | yes | Ma, J.+2020 |
| 2MASX J00253292 +6821442 | 52 | 246 | 30 | 24.1 | | 38.8 | Not detected | Ma, J.+2020 |
| NGC 3393 SBa | 53 | 257 | 438 | 24.3 | 10, 2600 | 40.9 | yes | Maksym+ 2017 Koss+2015 Jones+2021 |
| NGC4507 SABb | 57 | 275 | 140 | 23.6 | | 40.3 | | Matt+2004 Bianchi+2006 |
| MRK 3 S0 | 60 | 290 | 31 | 23.9 | 10, 2900 | 40.8 | | Bianchi+2006 NGC7212 |
| MRK 573 SAB0 | 72 | 349 | 114 | $>24.2^{n}$ | 9, 3100 | 40.8 | yes | Paggi+2012 Jones+2021 Bianchi+2010 Gonzalez-Martin +2010 |
| NGC 7212 | 115 | 550 | 148 | 24.1 | ~15, 8200 | 40.8 | yes | Jones+2020 |
| MRK 78 | 160 | 775 | 99 | 22.3 | 2.6, 2000 | 40.0 | Yes | Fornasini+ 2022 |
| 2MASX J04234080 +0408017 | 190 | 880 | 20 | 23.7 | 15, 13200 | 41.5 | | Fisher+2019 |

[a.] All morphological types are from the NED service (https://ned.ipac.caltech.edu/); [b.] in arcsec and parsec; [c.] luminosities in the 0.5-2 keV band; [d.] most extensive study listed first.

### 6.4.2 Chandra High Resolution Imaging Techniques

The Advanced CCD Imaging Spectrometer (ACIS) instrument on Chandra (Garmire et al., 2003) has several features that complicate the high-resolution imaging of AGNs. The back-illuminated ACIS-S3 chip (often simply called "ACIS-S") is normally used for imaging AGN extent as it had much better low energy effective area than the ACIS-I array at launch and is still ~40% higher below ~4 keV (Chandra Proposers' Observatory Guide[2], "POG", Section 6.5, Figure 6.5):

(1) The readout time for the CCD chips is 3.2 s for full-frame readouts. This leads to "pile-up" when a sufficiently bright source registers two or more photons in the same pixel in the same integration interval (POG Section 6.16), distorting the inner PSF by reducing the central count rate. A point source detected with 1 count/s, produces ~12% pile-up. This count rate corresponds to a flux of ~2 x $10^{-12}$ erg $cm^{-2}$ $s^{-1}$, and so includes the brightest few hundred AGNs. Using sub-frame readout at the 1/8 level (POG Section 6.13.1), with a 0.4 s readout time, reduces pile-up at that count rate to ~1% (POG Section 6.13.1). In 1/8 readout mode the field of view of ACIS-S is reduced to ~1.0 x 8.3 arcmin (POG Section 6.1, Table 6.2) from 8.3 x 8.3 arcmin in full frame mode (POG Section 6.13.1, Figure 6.17).

(2) The wings of the PSF on a few arcsecond scale are considerable compared to the relatively weak extended emission and must be corrected for. About 5% of the PSF lies between 1" and 2" radius from the center (POG Section 6.6, Figure 6.10).

(3) Sources with high count rates produce noticeable "trailed images" along the direction of the CCD readout (POG Section 6.13.1, Figure 6.18), further complicating the study of any extended emission near them along the readout direction.

(4) The ACIS soft response is much reduced from launch, most strongly since 2015, due to the accumulation of molecular contamination on the optical blocking filters (POG Section 6.5.1, Figure 6.7.)

To minimize pileup, nuclear PSF contamination, and trailed images, obscured and highly obscured AGNs are typically chosen, so that the nuclear photons are not directly visible at low energies (<3 keV). At these low energies, the extended emission is dominant (this chapter Section 6.4.3). The AGNs suited for these studies follow a careful selection process (this chapter Section 6.3 and, e.g., Fabbiano et al 2018a).

[2] https://cxc.harvard.edu/proposer/POG/html/index.html

Most importantly,

(5) The Chandra mirror ("high resolution mirror assembly", HRMA) has a sharp ~1/4 arcsecond central spike containing ~40% of the encircled energy, with ~20% within 0.1" (POG Section 4.2.3, Figure 4.6). Although the pixel size of the ACIS CCD imager is 0.492" (POG Section 6.1) this mismatch does not prevent sub-0.5" imaging with Chandra.

There are several ways to mitigate this mismatch:

(i) Complementing the ACIS observations with *High Resolution Camera (HRC) observations*, to image the innermost circumnuclear regions. The HRC has a smaller (0.13") pixels (POG Section 7.1). This method was used by Wang et al. (2009) in their study of NGC 4151, and Wang et al. (2012) in their study of the innermost regions of NGC 1068, which cannot be studied with ACIS, because of pileup. The HRC also has no pile-up or read-out streak, so that it can properly image even high-flux AGNs. The drawback to the HRC is that it had a lower effective area below ~1 keV, though ACIS contamination has lessened this disadvantage (POG Section 1.1.4). The HRC also has very limited energy resolution (E/ΔE ~ 1, POG Section 7.1).

(ii) Using *sub-pixel binning* to produce images from ACIS data. This method is made possible by the fact that the Chandra telescope scans over the source position in a well-known pattern of 16" amplitude Lissajous figures ('dither', see POG section 6.12). (The HST "multi-drizzle" technique approximates the same process, Koekemoer et al. 2003.) Dither allows the peak of the Chandra PSF to be sampled by sub-pixel binning down to 1/16 of a native ACIS pixel to show structures in high signal-to-noise images. This technique does not violate the Nyquist sampling theorem as the Chandra PSF is not a gaussian but has a sharp central spike, as noted above. The effectiveness and reliability of sub-pixel imaging with Chandra ACIS has been demonstrated by comparison with HRC images. Sub-pixel imaging with ACIS was applied to the ACIS data of NGC 4151 by Wang et al. (2011a), who demonstrated that it recovered the spatial information of the HRC observations of the same region (Wang et al. 2009). Moreover, the striking correspondence of the sub-pixel ACIS images with their HST and Atacama Large Millimeter Array (ALMA) counterparts gives extra confidence in the reliability of the method (e.g., Wang et al. 2011a; Fabbiano et al. 2018b; Feruglio et al. 2020). To illustrate the advantage of subpixel binning, Figure 6-2, compares the ACIS native image of the nuclear region of ESO 428-G014 (first panel on the left), with that obtained by binning the same photons in bins 1/16 of the instrument pixel (second panel). The improvement is more clearly seen in the third image where EMC2 image restoration has been applied, as explained next.

(iii) *Image processing* with adaptive smoothing and or image reconstruction (e.g., EMC2, Esch et al. 2004) can be used on sub-pixel images to enhance visually detected features. The third panel of Figure 6-2 shows as an example the 1/16 bin data processed with EMC2 image restoration (similar features are retrieved by using adaptive binning, see Fabbiano et al. 2018b); the last panel shows the EMC2 image plus 2-pixel Gaussian smoothing with the contours from the HST Hα image (Falcke et al. 1996). The close correspondence between Hα features and sub-pixel images is striking.

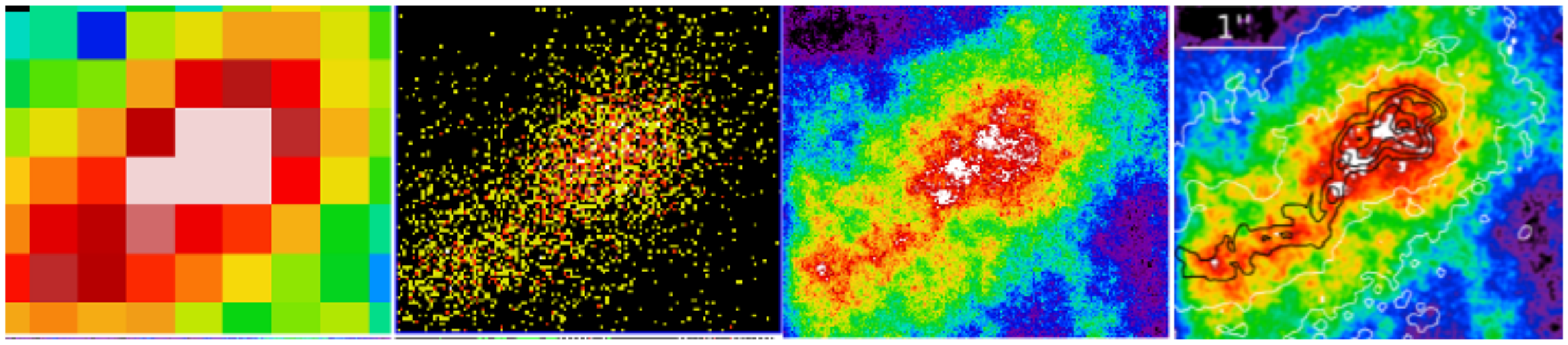

Figure 6-2: *First panel*: the Chandra ACIS image of ESO 428-G014 in the (0.3-3.0 keV) band; *Second panel*: same data with 1/16 binning; *Third panel*: EMC2 image restoration of the 1/16 binned image; *Fourth panel*: EMC2 image restoration smoothed with 2 image pixels with the HST Hα contours from Falcke et al. (1996). The Chandra images are adapted from Fabbiano et al. (2018b).

### 6.4.3 Broad-band (~0.3 – 2.5 keV) Soft X-ray Morphology

Figure 6-3 shows 25 CT AGNs that have detected soft X-ray extent mapped with Chandra. The CT AGNs are arranged in increasing distance as in Table 6-1 and demonstrate the existence of structures on scales from tens of parsec to several kiloparsecs. Most of the CT AGNs have a clearly favored geometry consisting of a two-sided elongated structure. The few one-sided structures (NGC4945, Circinus, NGC4388) are likely due to obscuration by a host galaxy dust lane (see references in Table 6-1). Within this general frame, there is considerable variation in the collimation of the structure and in the amount of sub-structure that is visible. To some extent this is due to the varying signal to noise of the observations, the distance to the sources, and to surface brightness limits attained.

The extended soft X-ray emission can be traced farther out by binning the data into radial profiles. Figure 6-4 shows this for ESO 428-G014 (Fabbiano et al., 2018a.) Extent is visible both in the major axis and perpendicular to this axis (cross-cone) out to a radius of about 15” (~1.7 kpc; see Section 6.6 for a discussion of the cross-cone extended emission and its implications).

Overwhelmingly, these soft-X-ray structures have a good overall alignment with the axis of extended narrow line region (ENLR) as traced by high [OIII]/Hα (or Hβ) ratios (Unger et al., 1987, Bianchi et al. 2006, and references in Table 6-1). MRK 573 is a particularly good example (Figure 6-5, Paggi et al., 2012).

These are radio-quiet AGNs, nevertheless they do have weak radio sources that appear like miniature double radio galaxies (Falcke et al. 1998; Nagar et al. 1999). The axes of these double radio sources are roughly, though not perfectly, aligned with the ENLR (Mundell et al., 1995, 2003), and so the soft X-ray structures also align with them. This correspondence between extended X-ray features and optical emission line regions is not limited to radio-faint AGNs but has also been observed in luminous radio galaxies (Torresi et al. 2009; Balmaverde et al. 2012).

The ratio of the optical [OIII] emission line to the soft X-ray flux is a well-known diagnostic of photoionization. Figure 6-6 shows that in NGC 4151 most of the points have a quite constant [OIII]/soft X-ray ratio from ~20 pc to ~1 kpc, implying a roughly constant ionization parameter. A wind provides a natural explanation for this constancy as the density of a wind drops as $1/r^2$, just as the ionizing nuclear flux does, so maintaining a constant ionization parameter. The strong correlation of the soft X-ray structures with those of the extended narrow line region (ENLR) and radio implies a direct connection to the central AGN for the soft X-ray extended emission. While the ENLR suggests a photoionization mechanism, the radio may point to shocks of the weak radio jet with the host ISM. Spectrally resolved imaging helps in disentangling these possibilities

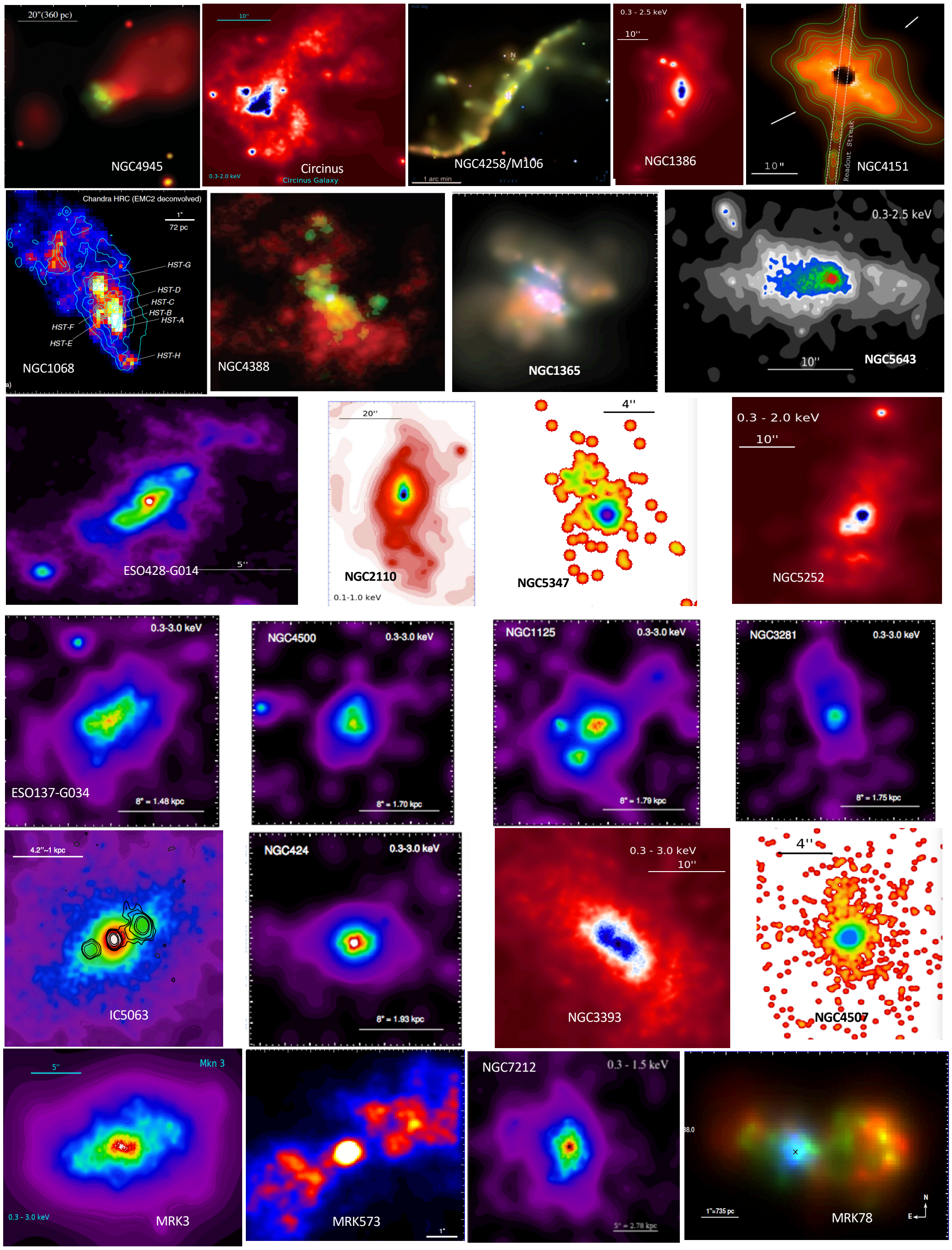

Figure 6-3: Chandra-imaged AGNs with extended soft X-ray emission (< 3 keV), also listed in Table 6-1. The images have different angular and physical distance scales. Note the (roughly vertical) ACIS readout streak in NGC4151, where there is a bright nuclear source. Color scales are either from the published papers (Table 6-1) or from reprocessing of Chandra data to visually enhance the soft features. This figure is meant for visual impact only. See the publications or use the Chandra data for quantitative measurements.

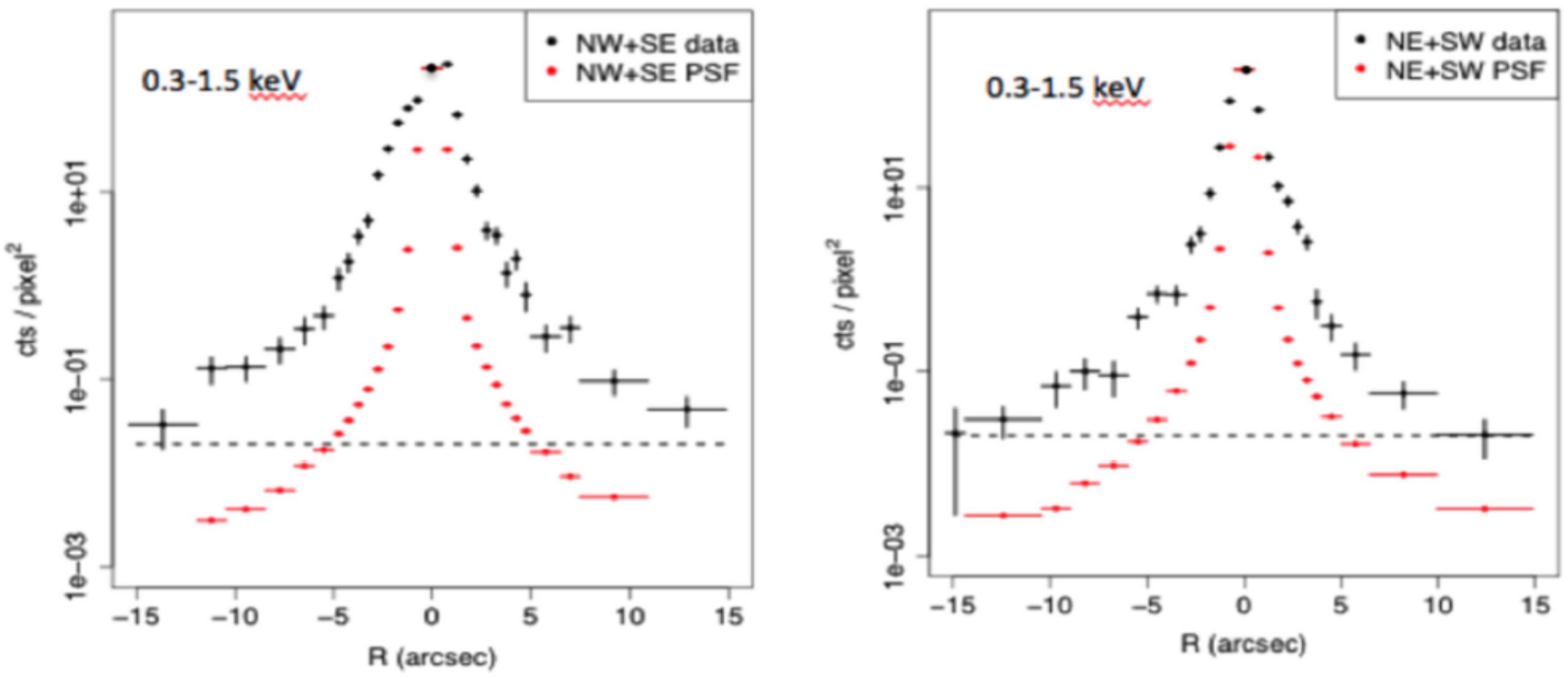

Figure 6-4: Radial profiles in the soft X-ray band (0.3 – 1.5 keV in this case) for ESO428-G014 parallel (left) and perpendicular (right) to the emission major axis (black data points, Fabbiano et al., 2018a; these data were background-subtracted.) The Chandra PSF is shown in red and is normalized to the observations at zero arcseconds. The horizontal dashed line is the background level. Clear excesses over the PSF are seen out to 15”.

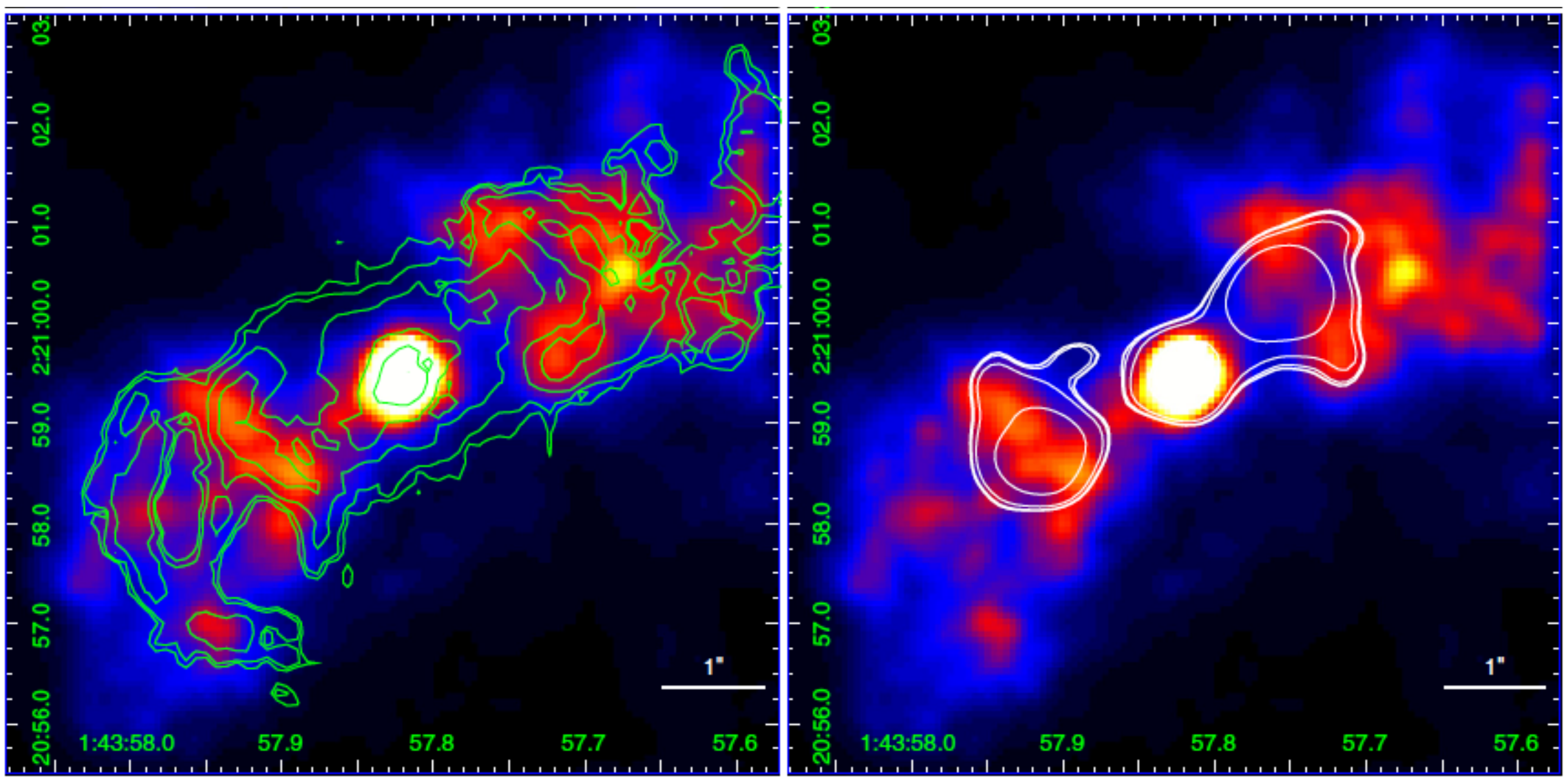

Figure 6-5: MRK 573 soft (0.3 – 2 keV) X-ray structure compared with, left, HST [OIII] and, right, VLA 6 cm maps (contours) (Paggi et al., 2012.)

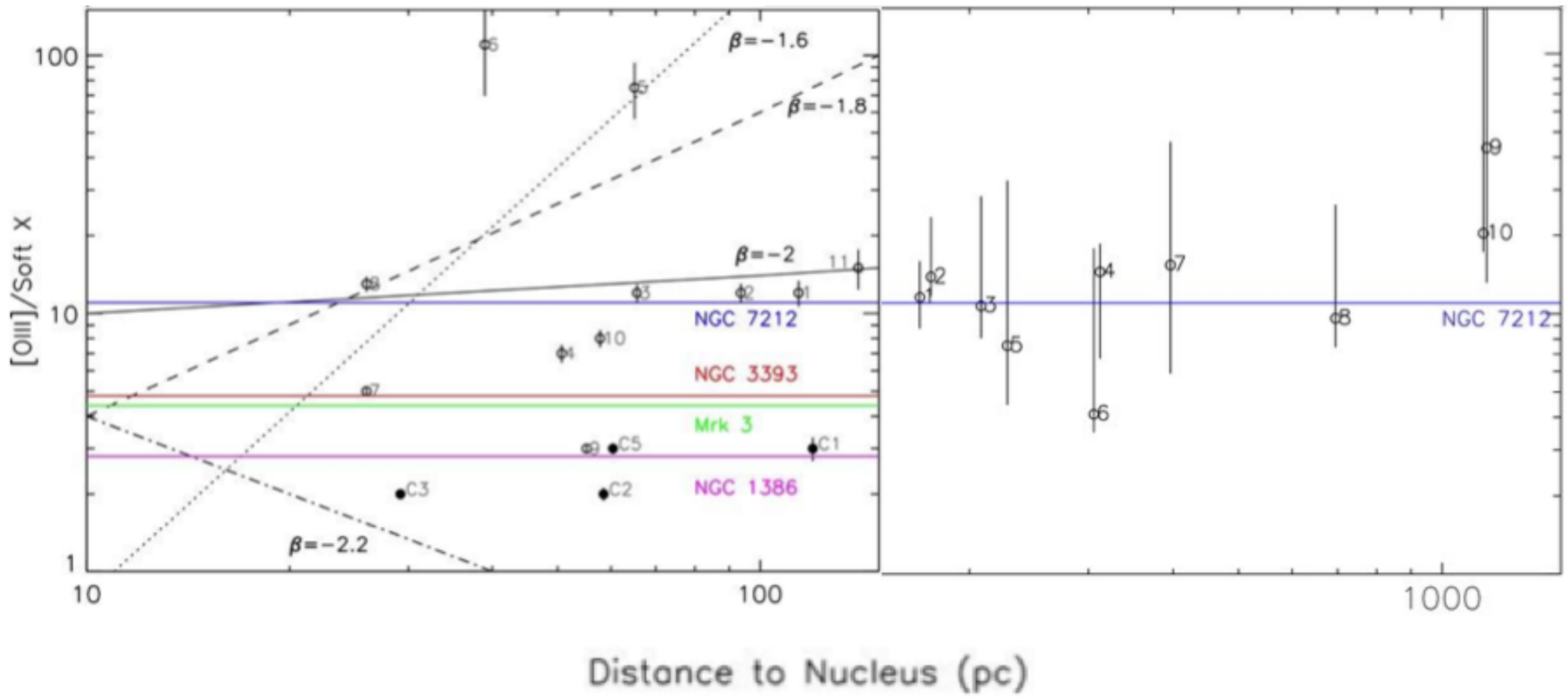

Figure 6-6: [OIII]/soft X-ray ratios for regions in NGC 4151 vs distance to the nucleus (pc) (adapted from Fabbiano and Elvis, 2019). The constancy of this ratio is suggestive of a wind model.

**NGC4258** is a special case, with much the most tightly collimated X-ray structure (Table 6-1, Figure 6-3, Wilson, Yang, & Cecil 2001, Yang & Wilson 2007). NGC4258 has unusually strong radio emission (at 20 cm) for a disk galaxy. The radio structure begins straight in the same direction as the VLBI jet that is perpendicular to the maser disk (Miyoshi et al., 1995, Herrnstein et al., 1995) and has almost N-S hot spots 0.84 kpc and 1.7 kpc from the nucleus. The hotspots are seen as Mach-shock-like arcs in H$\alpha$, and are detected in soft X-rays, but are too faint to extract spectra (Yang & Wilson 2007). They are an unusually clean example of radio jet/ISM interaction. Other radio structures follow the Hα "anomalous arms" (Cecil et al., 2000) extending out several arcminutes. The X-ray emission follows the anomalous arms more than the radio jet, though the radio hotspots are detected in X-rays. Wilson Yang, & Cecil (2001) propose a geometry in which the jet is pointed out of the galaxy disk plane and that in the inner ~350 pc the jet is cocooned by thermal X-ray gas, while further out the jet disrupts halo gas leading to that gas impacting and heating the galaxy disk, leading to the anomalous arms. The gaps in the disk HI emission in the anomalous arms supports this interpretation. No [OIII]/Hβ imaging exists to see if there is also a conventional ENLR, though Cecil et al. (2000) note a "nuclear loop" in Hα extending to ~230 pc North of the nucleus that may be an ENLR. The NuSTAR spectrum of the nucleus has not been published as of the time of writing.

### 6.4.4 Narrow-Band X-ray Emission Line Imaging

Chandra ACIS observations resemble integral field unit instruments such as MUSE on the VLTs (e.g., Venturi et al., 2021) by providing spectral information in every pixel. With good signal-to-noise (typically a few 100 counts) in the extended soft X-ray structures it is possible to make images in individual emission lines or blends. The primary lines and blends are labeled in Figure 6-7 (left) and listed in Table 6-2.

These lines and their ratios allow for diagnostics of the emission processes at work in these structures. Figure 6-7 (right) shows the Ne IX/O VII ratio map for the inner kpc of NGC 4151 (Wang et al. 2011.) Systematic factors of 2 changes in this ratio across the map are evident. Clearly conditions are not uniform across the soft X-ray structure and multiple emission mechanisms may be dominant in different locations. Fitting a single component model to the whole structure will lose much information. In particular, the Ne IX hot spots correspond to the end of the small radio jets in NGC 4151, and their emission can be explained with an additional thermal component from the shock-ionized interstellar medium (Wang et al. 2011b).

Table 6-2: Prominent emission lines in extended soft X-ray spectra of AGNs (Maskym et al. 2019, Paggi et al. 2012). Note that line blends are common at CCD energy resolution ($E/\Delta E \sim 10$ at 1 keV).

| **Line** | **Rest Energy (keV)** |
|---|---|
| CV He-$\gamma$ | 0.371 |
| C IV Ly-$\beta$ | 0.436 |
| N VII Ly-$\alpha$ | 0.500 |
| O VII f,i,r He triplet | 0.569 |
| O VIII Ly-$\alpha$ | 0.654 |
| Fe XVII | 0.720, 0.826 |
| Ne IX triplet | 0.915 |
| Ne X Ly-$\alpha$ | 1.022 |
| Ne IX He-$\gamma$ | 1.127 |
| Fe XXIV | 1.129, 1.168 |
| Mg X | 1.331 |
| Mg XII | 1.478, 1.745 |
| Si XIII | 1.894, 1.865 |
| Si XIV | 2.005 |
| S XV | 2.430 |

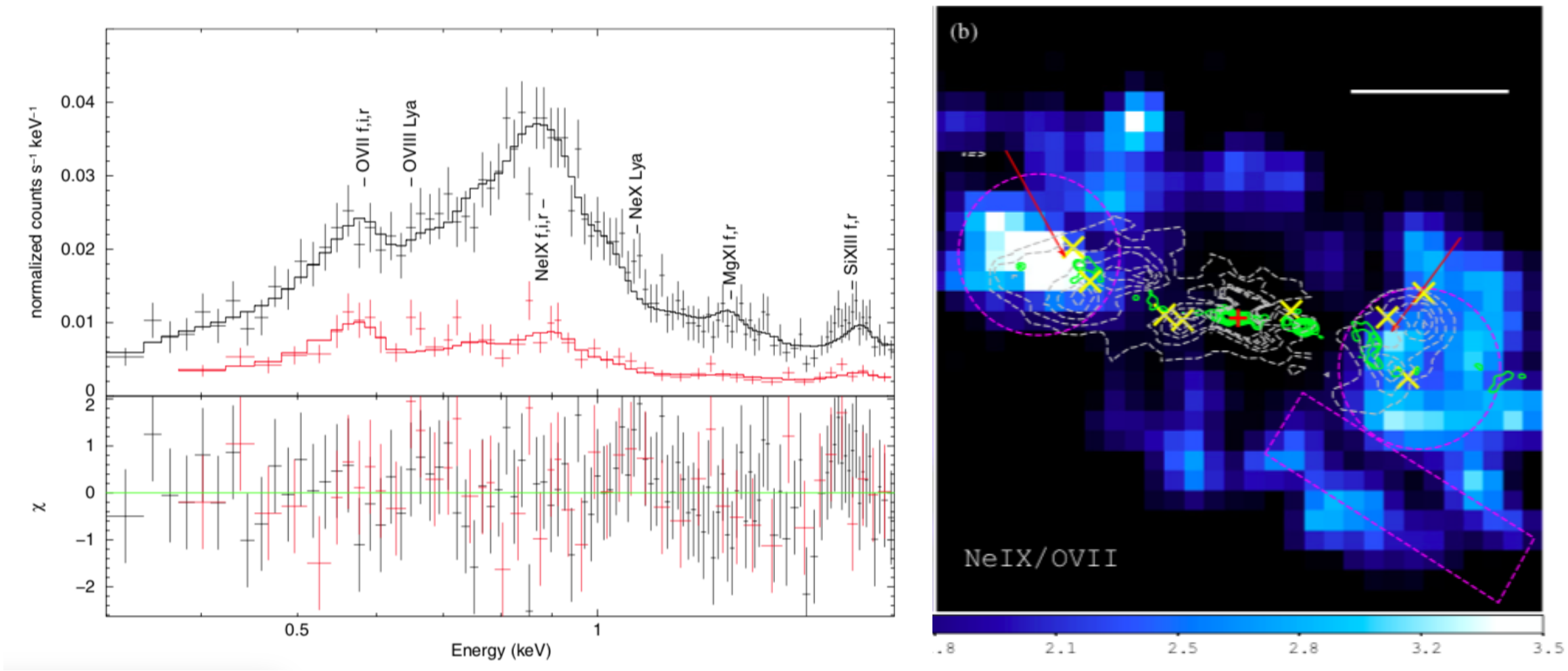

Figure 6-7: *Left,* Spectrum of the NGC 4151 hot spots, in black, from the extraction circles in the right panel, and of the unaffected area, in magenta, from the box in the right panel. The latter can be fitted with photoionization models, while the spectrum from the Ne IX excess areas require an additional thermal component; the best-fit residuals are shown in the bottom of the left panel (Wang et al 2011b, and references therein). *Right*, Ne IX /OVII ratio map from the ACIS observations of the nucleus of NGC 4151 (1/8 subpixel binning); green contours represent the radio jet, and dashed contour near-IR [Fe II] emission; yellow crosses mark the locations of the HST clouds with high velocity dispersion. The magenta circles are the extraction regions for the spectrum shown in black in the left panel, while the spectrum from the rectangular area is shown in magenta.

Figure 6-8 shows another example where comparison of emission in narrow bands corresponding to different X-ray emission lines provides localized constraints on the physical conditions of the emitting region. In this figure, the ratios between the emission in Si XIII, Ne (IX+X), and O VII from three different regions of ESO 428-G014 (Fabbiano et al. 2018b) are compared with a range of photoionization models, showing significantly different ionization parameters (U).

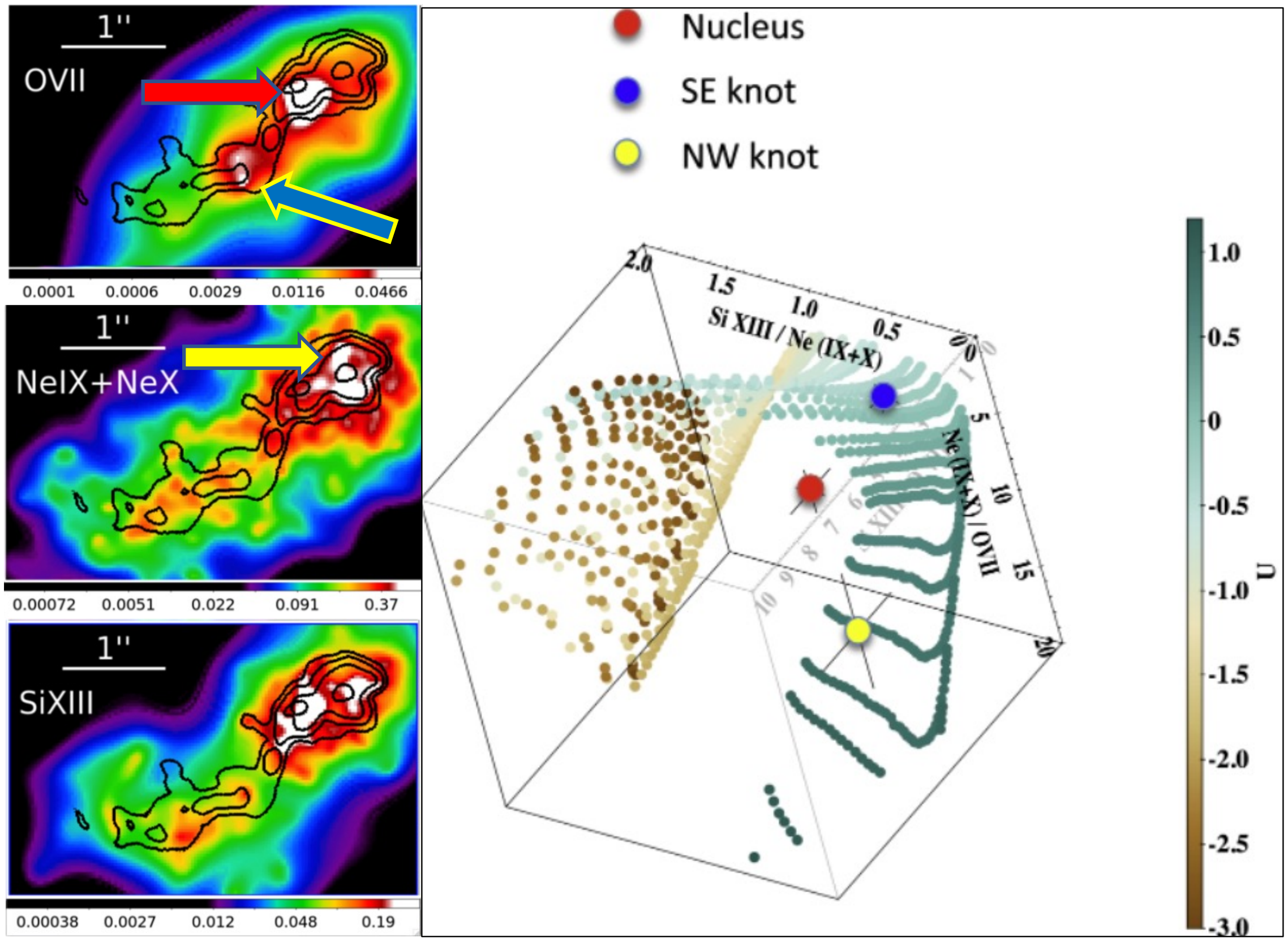

Figure 6-8: Adapted from Fabbiano et al. 2018b. *Left* - Chandra ACIS images of ESO 428-G014 in narrow bands including the emission lines marked. The colored arrows in the O VII and Ne (IX+X) maps point to the regions of interest for the diagnostic plots. Overlaid are H$\alpha$ contours from Falcke et al., 1998. *Right* - Diagnostic line ratios of three emission areas from the O VII, Ne (IX+X) and Si XIII images (nucleus, red; SE knot, blue; NW knot, yellow) plotted on photoionization model predictions for a range of log U (color scale). The three axes of the cube are Si XIII/Ne (IX+X), Ne (IX+X)/O VII, and in the background Si XIII/O VII. The measured ratios are plotted with their 1σ error bars. In the case of the SE knot the errors are the size of the point.

### 6.4.5 Spectra: Photoionization and Shock Excitation.

The strong line emission of the extended soft X-ray spectra of AGNs rules out, as dominant emission mechanisms, both synchrotron (such as seen in radio jets, Harris & Krawczynski 2006) and Compton scattering of the power-law nuclear spectrum by a polarizing screen of electrons (as in NGC 1068, Antonucci & Miller 1985). Two emission mechanism options remain: thermal gas, and gas photoionized by the central AGN. Optical line ratios show that the ENLR is photoionized (Osterbrock & Ferland 2006). However, the ENLR outflow speeds of ~300 - 1500 km $s^{-1}$ (Fischer et al. 2013) would, if shocked, produce gas temperatures, kT ~ 0.25 - 2 keV, giving emission in the soft X-ray band.

Deep exposures with Chandra/ACIS-S have given sufficient S/N to show that neither purely thermal nor purely photoionized gas alone is an adequate model (e.g., Wang et al. 2011c; Fabbiano et al. 2018a, Travascio et al. 2021). The fits cannot fully determine the components needed. Nevertheless, complex models consisting of at least 2 components of one mechanism are required in addition to at least one component of the other. Emission line imaging (Figures 6-7, 6-8) shows that conditions are clearly different in different locations in the extended structures.

There are common features: thermal components cluster around two temperatures, kT ~ 0.3 keV and ~1 keV (e.g., Fabbiano et al. 2018a); photoionized components cluster around two ionization parameters[3], log U ~ 1.5 (+/- 0.5) and -1.5 (+/- 0.75) (e. g., Travascio et al. 2021, Figure 6-9).

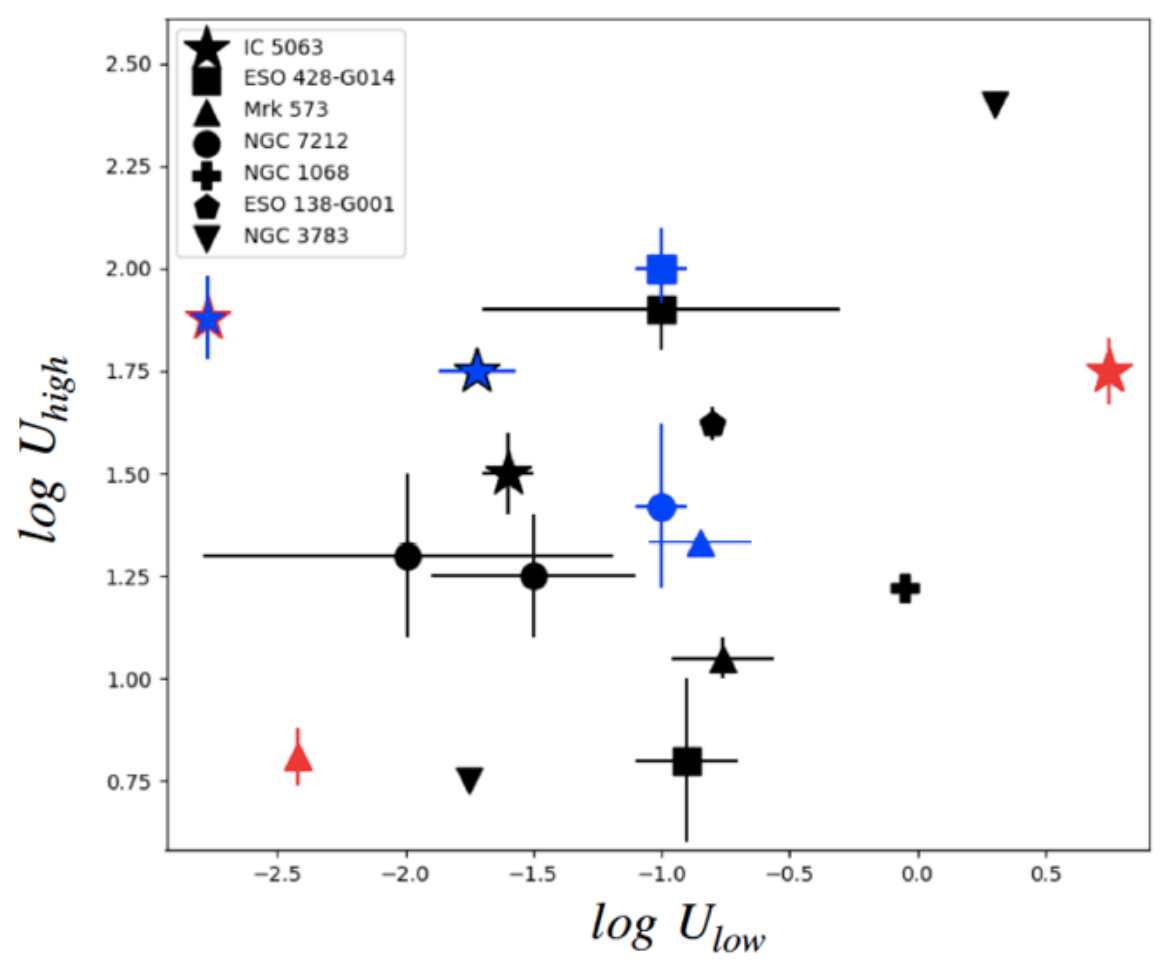

Figure 6-9: Distribution of photoionization parameters for high and low ionization components in the extended soft X-ray emission of 7 CT AGNs (Travascio et al. 2021).

---

[3] The ionization parameter U is the ratio of the incident ionizing radiation density to the electron density. I.e. U = Int{$L_\nu d_\nu$}/$4\pi r^2 c n_e$ from $\nu_R$ to infinity, where r is the distance of the gas from the source, $L_\nu$ is the ionizing luminosity, $\nu_R$ is the Rydberg frequency, and $n_e$ the electron density.

Thermal emission features tend to dominate in regions of strong interaction of the nuclear radio jets with the ISM (See Section 6.4.4; e.g., Wang et al. 2011b – Figure 6-7; Paggi et al. 2012; Travascio et al. 2021). The X-ray temperature then gives the speed of the shocks. Moreover, Chandra imaging in selected X-ray emission lines (O VII, O VIII, Ne IX-X, Si XIII) has shown that localized excesses of Ne IX-X emission may be associated with regions shock ionized by the impact of AGN jets (Figure 6-7, Wang et al. 2011b, Paggi et al 2012). Localized differences in these X-ray line images provide direct diagnostics of different physical properties (Fabbiano et al 2018b; Figure 6-8).

### 6.4.6 Seyfert and LINER emission co-existing in AGNs

Emission line galaxy nuclei may be AGNs, or starbursts, or LINERs (Kewley et al. 2001). While AGNs and starbursts have clear origins in photoionization by the active nuclear black hole or by young stars, respectively, the LINERs ("low ionization nuclear emission line regions", Heckman 1980) have long been ambiguous in their nature. Both shocks (Dopita & Sutherland 1995) and dilute AGN photoionization (Halpern & Steiner 1983) have been suggested as mechanisms to produce LINER spectra. LINERs tend to have low luminosities for AGNs, but a subset have clear broad components to their Hα lines pointing to the presence of an AGN broad line region (Ho 2008).

Several line ratio diagnostics have been used to classify LINERs (Veilleux & Osterbrock 1987, Baldwin, Phillips, and Terlevich 1981, "BPT"). The [OIII]/Hβ vs [SII]/Hα or vs [NII]/Hα plots are the most commonly used tool (Kauffman et al., 2003; Kewley et al. 2001, 2006) because their line pairs are close in wavelength and are minimally distorted by reddening (Veilleux & Osterbrock 1987). These plots are commonly called "BPT" diagrams. Classification of a galaxy as a LINER has mostly been based on nuclear spectra integrated over slit widths or fiber diameters of ~1 – 3 arcsec (3" in SDSS).

The advent of emission line imaging at high spatial resolution with HST and with ground-based integral field units (IFUs) has made possible "excitation imaging" of individual nuclei (Ho et al., 2014; Cresci et al. 2015; Davies et al. 2016). These studies have begun to show that even nuclei classed as AGN often contain regions that are LINER-dominated (Cresci et al. 2015; Maksym et al. 2016; Ma et al. 2021). NGC3393 is a striking example. A BPT diagram of every pixel in continuum subtracted HST emission line images shows a wide spread of excitation (Figure 6-10, left). When these are divided into AGN and LINER types and are mapped back onto the sky a clear pattern is seen (Figure 6-10, right): the AGN regions (red) are mostly correlated with the intense soft X-ray emission features, supporting a prevalent ISM photoionized by incident nuclear photons. Instead, the LINER emission regions (yellow) surround the AGN emission in a "cocoon" (Maksym et al. 2016).

A survey of seven similar AGNs shows that LINER regions are common in CT AGNs and that they often cocoon the AGN emission (Ma et al. 2021). This geometry suggests a shock origin for LINERs as the X-ray wind expands into the ISM, though not conclusively. It definitely highlights that the interaction of the active nucleus with the interstellar medium is complex. As we have discussed in Sections 6.3.3 and 6.3.4 this complexity is not limited to the LINER cocoons but is also found in the extended AGN emission. ACIS spectra of the extended AGN-dominated regions require complex multi-component models spanning a range of thermal emission temperatures and AGN photoionization levels (Wang et al. 2011b, c; Fabbiano et al. 2018a; Travascio et al. 2021).

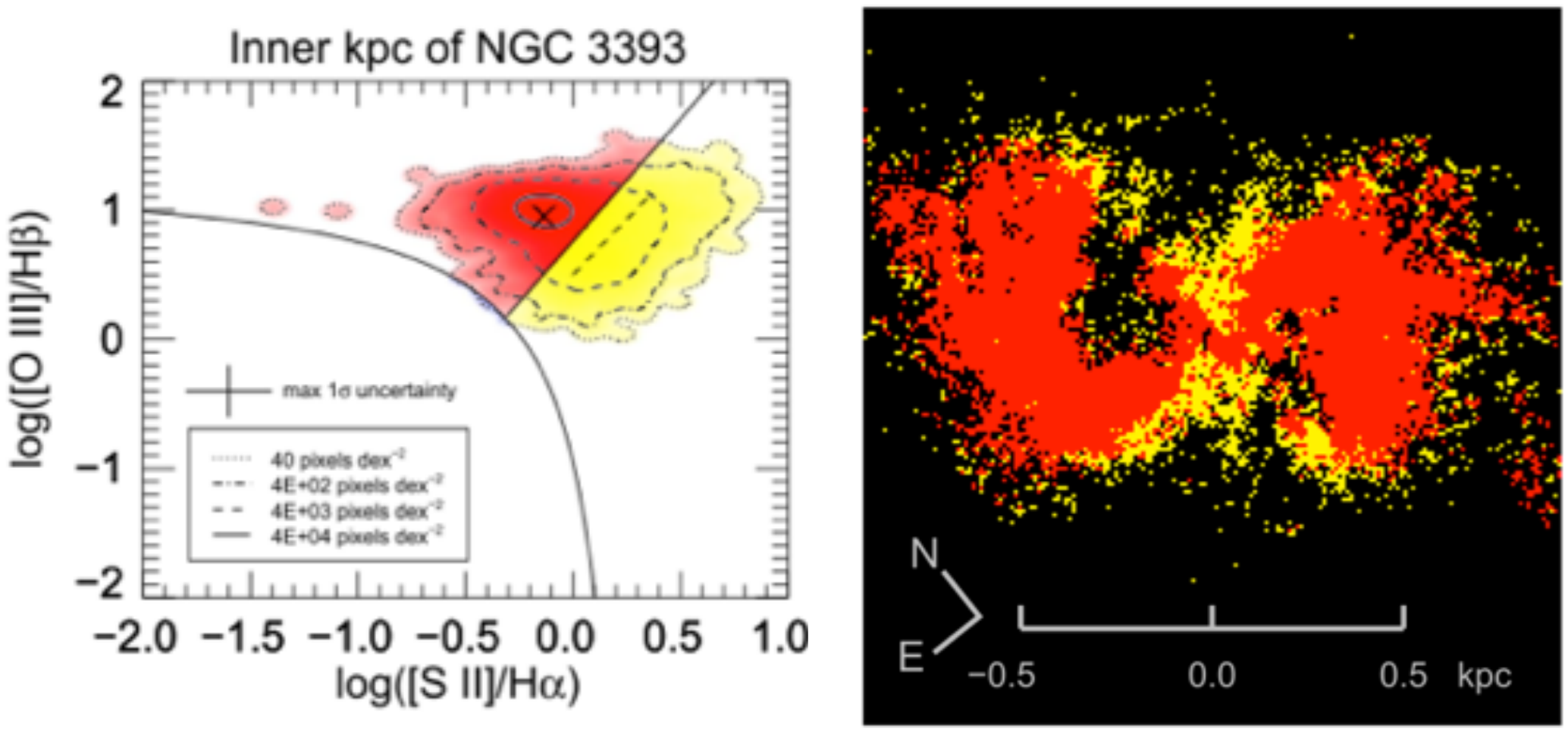

Figure 6-10: The AGN NGC 3393 (Maksym et al. 2016). *Left*: Pixel data in a BPT diagram. *Right***:** The same data in a sky-map (yellow= LINER; Red=Seyfert). The higher excitation 'Seyfert' regions are associated with more intense X-ray emission in the Chandra image (Maksym et al. 2017). High angular resolution data are needed to resolve these regions, which would not be distinguishable in most ground-based telescope data (the SDSS fiber is ~0.75 kpc diameter on NGC 3993.)

## 6.5 CHANDRA IMAGING: DISCOVERY OF EXTENDED HARD CONTINUUM AND Fe Kα (NEUTRAL)

In the standard AGN model (see Section 6.2.3), the hard continuum (approximately energies greater than 2.5 keV) and the 6.4 keV Fe Kα line originate from the interaction of the AGN photons with the obscuring torus and should be confined to the nuclear region (see Section 6.3.1). The hard continuum is thought to arise from the reflection of hard photons from the AGN on the dense clouds of the torus, while the 6.4 keV Fe Kα line would be due to fluorescence excited by the nuclear photons (Section 6.4.2). Since the obscuring torus is estimated to have diameters of a few parsecs (e.g., Gandhi et al. 2015), this emission should appear point-like even with the Chandra telescope (resolution ~0''.3, Section 6.4.2).

Chandra observations have instead shown that there are extended components of the hard continuum and of the 6.4 keV neutral Fe Kα line. This extended emission was first reported in NGC 4945 (Marinucci et al. 2012), a galaxy at only ~3.7 Mpc distance (1'' ~ 18 pc), where a ~200 pc feature was detected in the direction of the torus (perpendicular to the bicone). In two other nearby CT AGNs also observed with Chandra, the Circinus Galaxy (Arevalo et al. 2014) and NGC 1068 (Bauer et al. 2015), extended hard continuum and 6.4 keV Fe Kα line components were found, instead, in the direction of the bicone (∼600 pc in Circinus, >140 pc in NGC 1068). Deep Chandra imaging of the CT AGN ESO 428-G014 (Fabbiano et al. 2017, 2018a; Figure 6-11) for the first time showed hard continuum and Fe Kα emission on a comparable kiloparsec-scale size as the soft X-ray extended emission along the bicone direction.

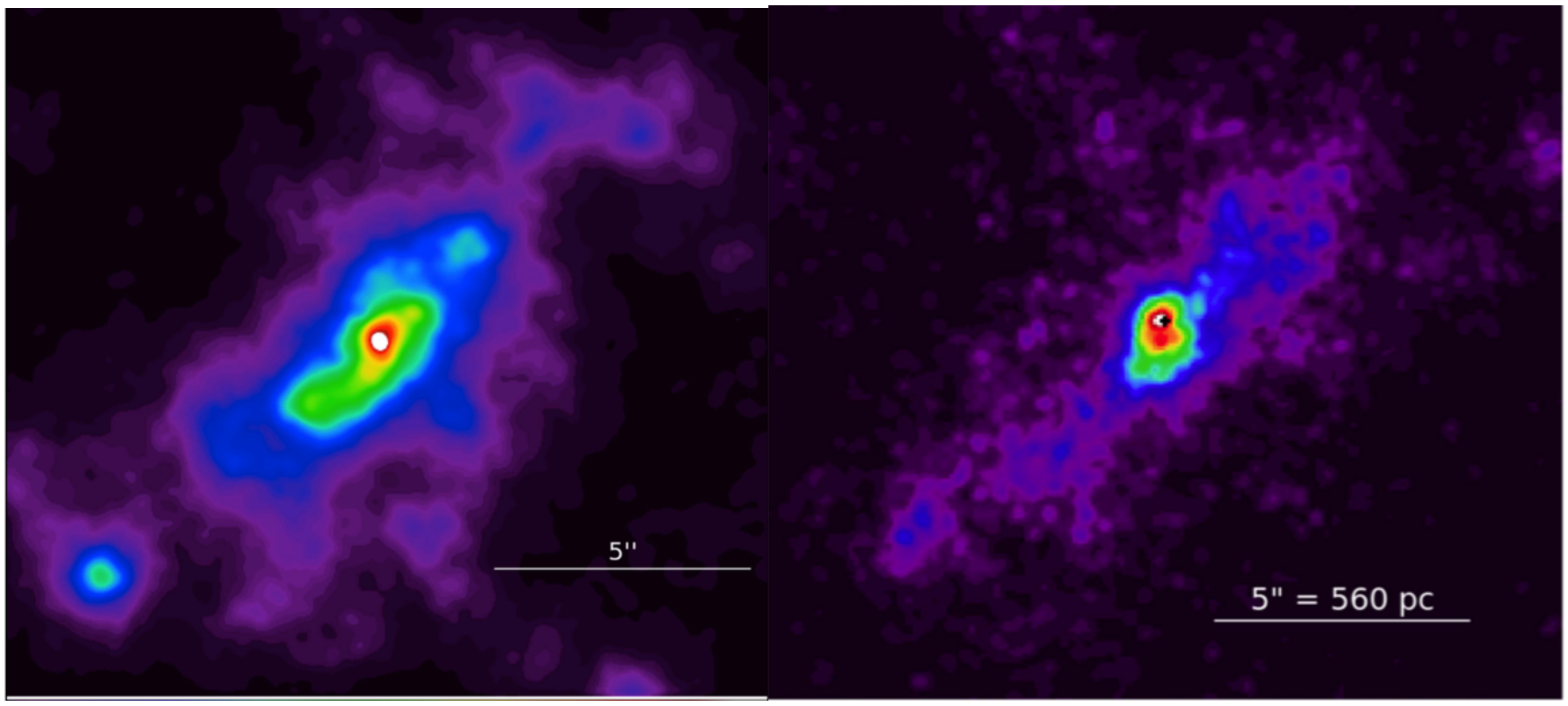

Figure 6-11: *Left,* 3-6 keV continuum image of ESO 428-G014 (Fabbiano et al. 2017). *Right,* Fe Kα image (Fabbiano et al. 2018a).

These discoveries were followed up by imaging studies of several nearby CT AGNs with Chandra, establishing that the large-scale diffuse hard continuum and 6.4 keV Fe Kα line emission are a general feature of these systems (see Table 6-1; Fabbiano et al. 1918a, b; Jones et al. 2020, 2021; Ma et al. 2020; Travascio et al. 2021; Fornasini et al. 2021).

Could the hard excess be due to X-ray binaries in the galaxy rather than being related to AGN activity? In ESO428-G014 the luminosity of the hard component at radii 1.5"-8" from the nuclear centroid is estimated to be $L_X$(3-7 keV) ~ 1.3 x $10^{40}$ erg $s^{-1}$ (Fabbiano et al. 2018a), based on best-fit model assumptions. For comparison, the line-dominated soft luminosity is $L_X$(0.3-3 keV) ~ 1 x $10^{40}$ erg $s^{-1}$. The hard emission cannot be explained with the integrated contribution of X-ray binaries in the region: both the luminosity and the spectral characteristics are not consistent with the expected X-ray binary contribution (Fabbiano et al. 2017). Moreover, the spatial distribution of the hard emission, which follows that of the soft X-ray component and the optical line emission bicone, indicates that this emission is related to the AGN irradiation of the ISM. Similar conclusions can be reached from the study of the other deeply imaged AGNs (see references above). More recently, a compilation of a sample of CT AGNs with shorter less-sensitive Chandra observation has shown extended soft and hard emission in excess of what their populations of XRBs would produce, even at ten or more times lower luminosities (Figure 6-12; Ma, Elvis, et al. 2022, in preparation). Only in a few cases is an X-ray binary origin not ruled out.

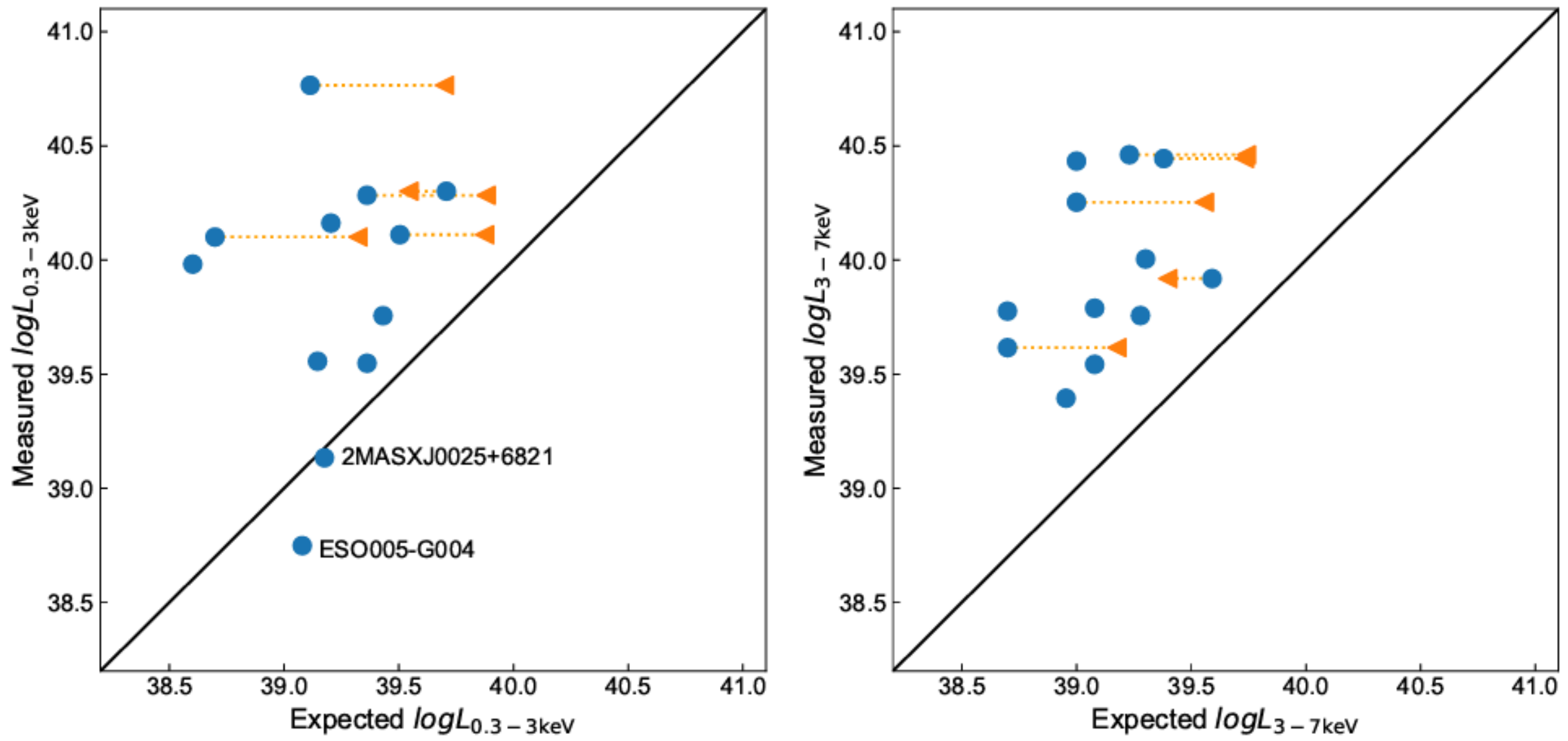

Figure 6-12: Measured extended X-ray luminosity versus expected luminosity from XRBs for the 0.3-3.0 keV (Left) and 3.0-7.0 keV bands (Right). The blue circles are the expected X-ray luminosities from LMXBs. The orange triangles are the expected X-ray luminosities from HMXBs or mixed populations. The diagonal line denotes the 1:1 ratio.

These results suggest that hard photons from the AGN, escaping along the same torus axis as the soft photons that give rise to the bicone, are then scattered / reflected via fluorescence by dense molecular clouds in the disk of the host galaxy, similar to what is seen in the Milky Way Galactic Center (Koyama et al 1996). In ESO 428-G014, >30% of the observed hard continuum is not associated with the nuclear source (Fabbiano et al. 2018a). This raises a challenge for the constraints posed on the AGN from spectral modelling based on the standard model.

Compton scattering off ISM clouds requires dense clouds to scatter the radiation efficiently given the electron scattering cross-section (Fabbiano et al. 2017). An energy dependence in the extent of the emission was found in ESO 428-G014, with the softer (lower-energy) component being more spatially extended than the harder emission (Fabbiano et al. 2018a). As noted in Fabbiano et al. (2018a), this energy dependence suggests that the dense clouds responsible for the scattering of the hard AGN photons are more concentrated at smaller galactocentric radii, as in the Milky Way (e.g., Nakanishi & Sofue 2006). These high-density clouds must be clumped so that the soft ionizing X-rays from the AGN can escape to larger radii. This result was confirmed by other observations of CT AGNs with Chandra, although with some differences in each case that may be related to the relative orientations of the ionization cones and the molecular cloud distributions (Ma et al. 2020; Jones et al. 2021; Travascio et al. 2021). Figure 6-13 shows this effect for the bicone emission of a sample of five CT AGNs studied by Jones et al. 2021.

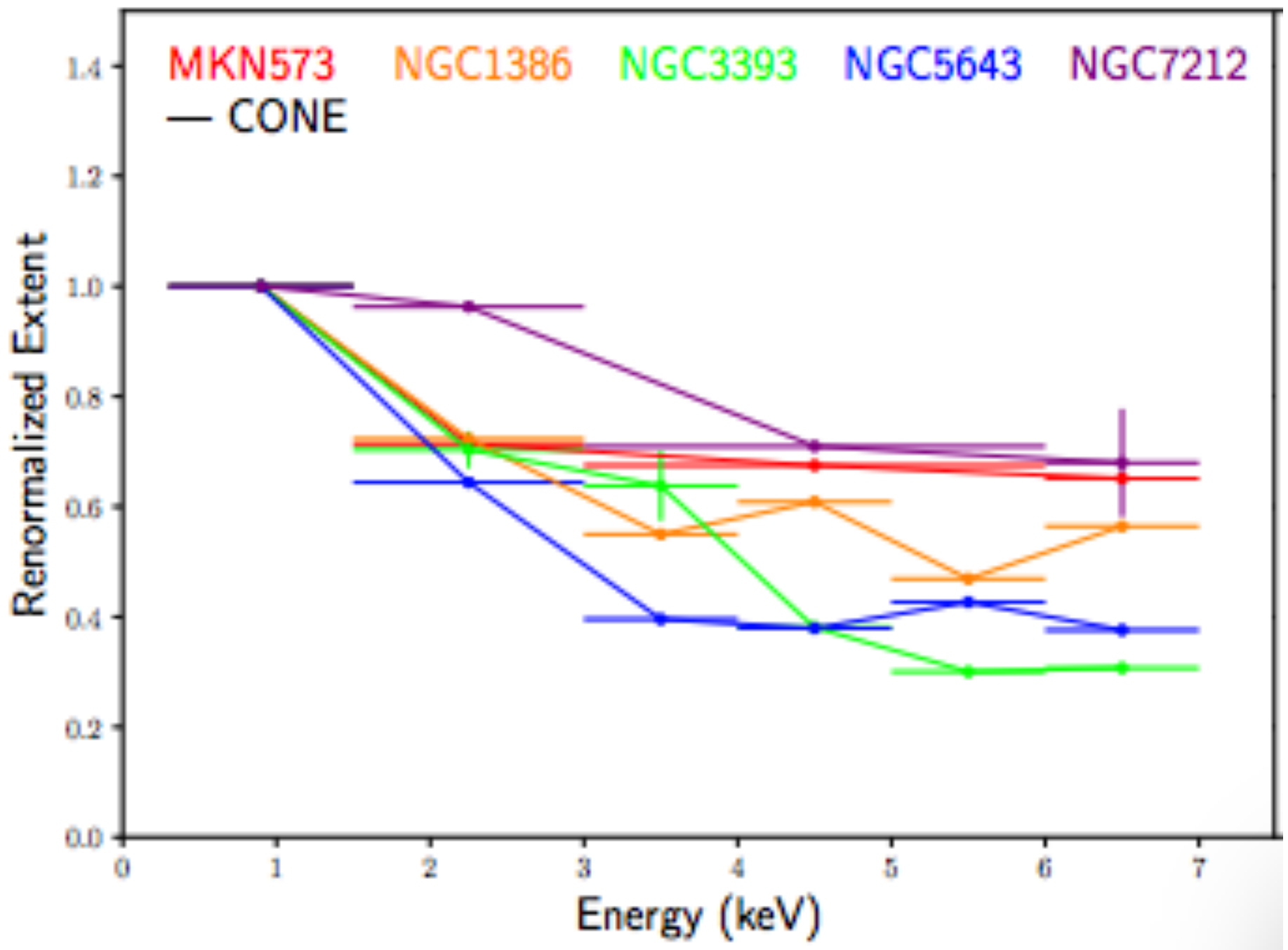

Figure 6-13: Renormalized extent (kpc) of the bicone at 1% of the peak surface brightness for the sample of five CT AGNs as a function of energy (from Jones et al. 2021).

### 6.5.1 The effect of fast shocks: the Fe XXV Kα line emission

Besides the prominent neutral Fe Kα line at 6.4 keV, ionized Fe Kα lines at emitted energies of 6.7 keV (Fe XXV) and 6.9 keV (Fe XXVI) have been detected in CT AGNs. In the case of the dual AGN merging galaxy NGC 6240 (z=0.02448 +/- 0.00003; Downes et al. 1993), these ionized lines, and the hard continuum emission also present in the spectrum of NGC 6240, are consistent with thermal emission from a shock-heated gas of at least kT ~ 6 keV, corresponding to a fast shock velocity of ~ 2000 km $s^{-1}$. Figure 6-14 shows the ACIS spectrum extracted from the central region of the NGC 6240, where Fe XXV and Fe XXVI are quite strong (Wang et al. 2014). These shocks may be due to nuclear winds from the AGN interacting with the gaseous interstellar medium or to winds originating from intense star formation in the inner regions of NGC 6240 (Wang et al. 2014; Feruglio et al. 2013).

Localized regions of Fe XXV and Fe XXVI emission have been reported in the ~200 pc nuclear flattened feature in NGC 4945, where it may be caused by photoionization by the intense nuclear photon field (Marinucci et al. 2017, see Section 6.8) and in IC 5063, from the region where the radio jet appears to terminate, possibly because of interaction with a cloud in the galaxy (Travascio et al. 2021).

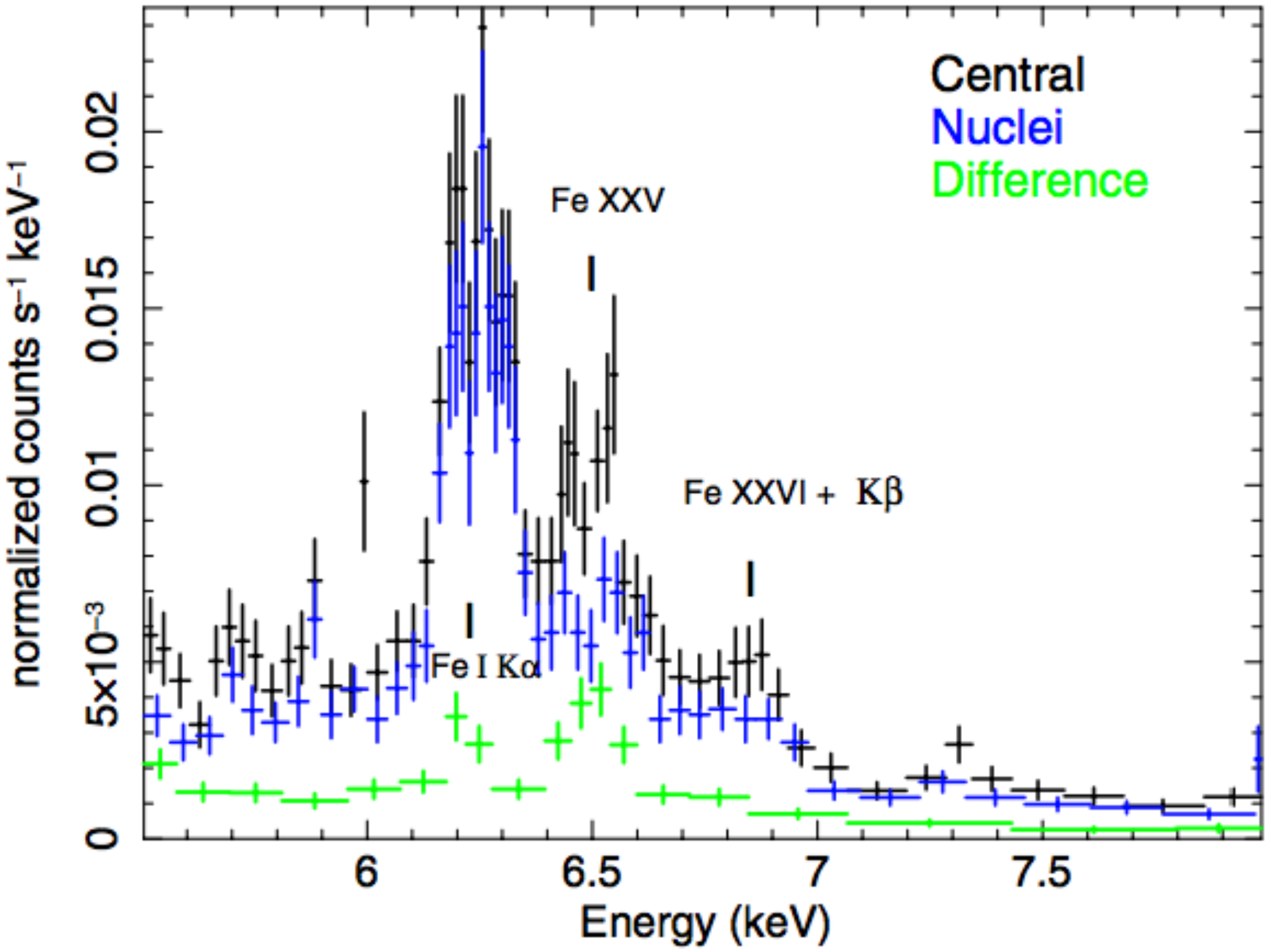

Figure 6-14: ACIS spectra extracted from the central 10'' radius region of NGC 6240 (black), the central 2'' radius region including the nuclei (blue) and the circumnuclear region, which excludes the nuclei (green). The neutral Fe K$\alpha$ line is strongest in the nuclear spectrum, while Fe XXV is relatively more prominent at larger radii (Wang et al. 2014). The spectrum is displayed as observed (i.e., it is affected by the redshift of the host galaxy, z=0.02448).

### 6.6 CROSS-CONE EMISSION – LEAKY TORUS OR JET-STIMULATED OUTFLOWS?

In the standard model (Section 6.2.3), AGN photons are collimated by the torus and escape only along the axis of the torus. This is consistent with the detection of bicones in optical emission lines and soft X-rays emission (see Sections 6.2.2, 6.4.3). In the standard model, there should be no emission detected in the direction perpendicular to the axis of the ionization / wind cone, the cross-cone direction. However, recent Chandra observations of nearby AGNs have shown that cross-cone extended X-ray emission is instead present. It has been reported in NGC 4151 (Wang et al. 2011c), MRK 573 (Paggi et al. 2012), ESO 428-G014 (Fabbiano et al. 2018a), IC 5063 (Travascio et al. 2021), and other AGNs (see Jones et al 2021).

The extent of the cross-cone emission at ~1 keV ranges from radii of ~4 kpc down to ~50-100 pc, with the most extended soft cross-cone components being reported for IC 5063 (r ~4 kpc, Travascio et al 2021), ESO 428-G014 (r ~1 kpc; Fabbiano et al 2018a), NGC 7212 and NGC 3303 (r

~ 1kpc and ~700 pc respectively, Jones et al. 2021). At higher energies the extent is typically smaller, down to ~ a few 100 pc or less (Jones et al. 2021). The most extended hard cross-cone components are reported in ESO 428 -G014 (r~400 pc, Fabbiano et al 2018a), and NGC 7212 (r ~ 700 pc, Jones et al. 2021).

This cross-cone emission signals a departure from the simplest standard model assumptions, and its explanation is not clear-cut: (a) It may be due to a porous cloud structure in the obscuring structure (Nenkova et al., 2002, 2008). The possibility of a porous obscuring cloud structure is supported by the absorbing column density variability in many AGNs (Torricelli-Ciamponi et al. 2014). It is particularly well-documented in two AGNs, NGC 1365 (Risaliti et al. 2005) and NGC 4945 (Marinucci et al. 2012), which demonstrates that the obscuring structure is composed of discrete clouds with distinct boundaries that at times move out of the AGN line of sight. However, their time variability implies cloud velocities consistent with the broad-line region, too fast for these clouds to be in the torus (Risaliti et al. 2007). Several slower moving clouds that are more consistent with a location in the torus were found by Markowitz, Krumpe, and Nikutta (2014), but these are quite rare. Only 4 events at $>10^5$ $R_g$ were found in 17 years of Rossi X-ray Timing Explorer (RXTE) data on 45 AGNs. (b) Alternatively, cross-cone X-rays may be the result of the interaction of a nuclear radio jet with the dense interstellar medium of the galaxy disk within which this jet is propagating (Mukherjee et al. 2018). (c) Lastly, vestigial emission from a past era of more intense AGN activity may also be a possibility (e.g., Wang et al. 2010; see Section 6.7). Below, we discuss the observational results, and the constraints they pose.

The cross-cone emission was first discovered in the Chandra observations of NGC 4151 (Wang et al 2011c), and soon after in MRK 573 (Paggi et al. 2012), extending outwards from the nuclear region. In both cases, these authors advocate leakage from the obscuring torus. This interpretation is supported in the case of NGC 4151 by peculiar emission line kinematics that suggest clouds outflowing in the cross-cone direction, which would need to be ionized by escaping nuclear photons (Das et al. 2005). In both NGC 4151 and MRK 573, the cross-cone X-ray emission can be fitted with two-component photoionization models, and in both cases the lower ionization component suggests a lower ionizing flux received by the clouds than in the cone direction. Wang et al. (2011c) suggest that this effect may be due to the nuclear continuum filtered by warm absorbers (Krongold et al. 2007) or the walls of the ionization cone (Kraemer et al. 2008) and estimate that about 1% of the nuclear ionizing flux is seen by clouds in the cross-cone direction. Similarly, Paggi et al. (2012) derived an upper limit of <3% for the cross-cone ionizing flux. In NGC 4151 the cross-cone emission is enveloped in the kpc-size hot bubble surrounding the nucleus (Wang et al. 2010; see Section 6.7), however both the higher intensity near the nucleus and the presence of outflowing clouds in that direction are consistent with a continuing outflow.

The deep Chandra exposures on ESO 428-G014 (Fabbiano et al. 2018a) and IC 5063 (Travascio et al. 2021) allow a good characterization of the diffuse cross-cone X-ray emission in these two AGNs. In the ‘torus leakage scenario’, in which the torus is porous, the opening angle of the bicone and the ratio of the X-ray emission detected in the cross-cone and bicone regions can be used to evaluate the cross-cone transmission of the torus. In ESO 428-G014 the transmission would be 10%, while in IC 5063 would be double that, 20%.

Jones et al. (2021) perform a comparison of the spatial properties of the extended X-ray emission in a sample of 5 CT AGNs (MRK 573, NGC 1386, NGC 3393, NGC 5643, and NGC 7212), for which high-resolution Chandra observations are available. Extended emission is detected in several AGNs at high energies, up to 8 keV. This study finds a range of energy-dependence of the extent of the cross-cone emission, from no energy dependence in most cases, to a strong energy dependence in NGC 3393 (similar to that observed in ESO 428-G014 and IC 5063, see Fabbiano et al 2018a; Travascio et al. 2021). If this emission is due to AGN photons escaping from a porous torus, these results may suggest different scale heights of the dense reflecting clouds in the different galaxies, (see Fabbiano et al. 2017).

In both ESO 428-G014 and IC 5063, the cross-cone emission is more prominent at energies < 1.5 keV, and the overall aspect ratio of the emission is rounder than at the higher energies (Fabbiano et al. 2018a). This shape suggests that this soft emission in ESO 428-G014 may be due to a quasi-spherical thermalized soft gaseous component, with a temperature of ~0.7 keV. In IC 5063 the cross-cone extended emission is also prevalently soft and, within statistics, has an energy dependence similar to that of ESO 428-G014. If this emission is thermal, its temperature is between 0.3 and 1.4 keV (Travascio et al. 2021). These soft extended components could be due to thermalized hot halos trapped in the gravitational potential of the galaxy (see Bogdan and Vogelsburger, Chapter 4). However, both the temperature of the emission and the roundness of the low-energy extended emission in ESO 428-G014 and IC 5063 (Figure 6-15) are also consistent with the results of 3D relativistic hydrodynamic simulations by Mukherjee et al. (2018) of the interaction of radio jets with a dense molecular disk, tailored to IC 5063 (see Fabbiano et al. 2018b; Travascio et al. 2021).

In these simulations, outflows from the interaction region in the galaxy disk produce a hot cocoon enveloping the interaction region. Since the CT AGNs where cross-cone extent has been reported so far all host small radio jets, this model offers an alternative (or additional) explanation for the cross-cone emission. Additional support for galaxy disk outflows - that would be consistent with the jet-disk interaction simulations - is provided by recent narrow band filter observations with the HST of BPT diagnostic forbidden ([O III], [S II]) and Balmer (Hα, Hβ) lines in the NLR of IC 5063 (Maksym et al. 2021), leading to the discovery of a 700 pc diameter giant loop of ionized gas emerging from the disk. These authors speculate that large-scale lateral outflows may also be responsible for the ~10 kpc optical 'dark rays' also visible in this galaxy ("crepuscular rays", Maksym et al. 2020). Moreover, Venturi et al. (2021) have detected [OIII] turbulent motions in the cross-cone direction of IC 5063 and three other AGNs with small radio jets (NGC 5643, NGC 1068, NGC 1386), suggesting that the cross-cone X-ray emission may be generally associated with turbulent cross-cone outflows.

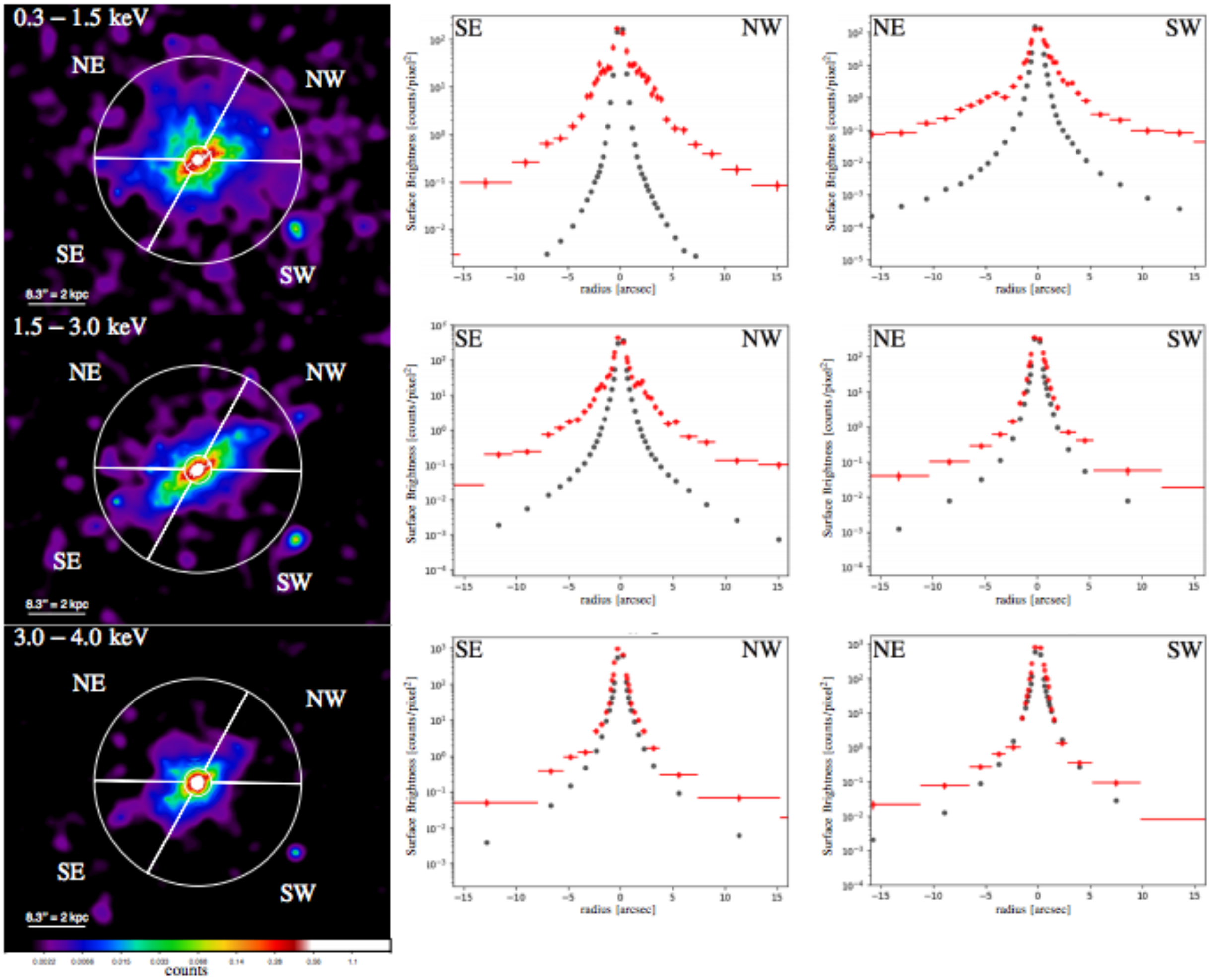

Figure 6-15: Images of IC 5063 in three energy bands: 0.3-1.5 keV (top), 1.5-3.0 keV (middle) and 3.0-4.0 keV (bottom). The radial profiles extracted in the bi-cone and cross-cone direction are shown next to each image in red; the black points represent the Chandra point response function in the same energy band (Reproduced from Travascio et al. 2021).

In the softer energy band (<1.5 keV), both the extent of the emission and evidence of outflows in the optical emission lines are in agreement with the scenario of outflows stimulated by the interaction of the radio jets in these galaxies with the dense interstellar medium. However, the presence of hard continuum and neutral Fe Kα line emission in these cross-cone regions in several CT AGNs (e.g., Jones et al. 2021) also may suggest the escape of hard nuclear photons from a porous obscuring region. The spatial extent of the cross-cone component is smaller at the higher energies (typically a few 100 pc or less), so the scattering clouds could be part of a thick disk structure of the host galaxy (such as that seen in the Milky Way, Gilmore & Reid 1983). But in some case the escaping photons could be scattered by (and in some cases cause fluorescent Fe

Kα emission in) clouds above the host galaxy disk that may perhaps have been displaced to these regions by jet stimulated galaxy outflows. Finally, we cannot exclude some contribution from unresolved X-ray binary emission in the host galaxy in some of the lower luminosity cases (Figure 6-12.)

## 6.7 X-RAY IRRADIATION OF MOLECULAR CLOUDS IN THE CENTRAL ~100 PC: IMAGING THE TORUS AND AGN FEEDBACK

The 'standard AGN model' (Section 6.2.3) posits that both the hard (>2.5 keV) continuum and the 6.4 keV Fe Kα line originate from the interaction of nuclear X-rays interacting with an obscuring nuclear torus. IR observations of AGNs suggest that this torus may have a diameter on a few pc-scale (e.g., Gandhi et al. 2015). While this diameter is too small for imaging with Chandra, there is a growing body of work indicating the presence of extended, flattened, nuclear structures on somewhat larger scales. In one case at least, NGC 4151, an absorbed feature of ~600 pc extent is consistent with obscuration from cold ISM in front of the nuclear region. The shape of this feature follows CO and absorbed optical emission, suggesting the presence of an extended circumnuclear disk (Wang et al. 2011a).

Flattened nuclear structures emitting hard X-ray continuum and Fe Kα have also been observed. In nearby AGNs the availability of both *ALMA* and *Chandra* allow the direct imaging of both the large-scale structure enveloping the nuclear obscuring 'torus' and the result of the nuclear X-rays interacting with this structure.

The first indication of a flattened hard (>2 keV) X-ray emitting central structure was found in the high-resolution *Chandra* ACIS images of CT AGN NGC 4945 (Marinucci et al. 2012), a galaxy at a distance of only ~3.7 Mpc (1'' ~18 pc). This structure is perpendicular to the extended soft X-ray bicone, in the direction expected for the nuclear torus, and has a diameter of approximately 200 pc. This diameter is larger than expected for the nuclear torus, so the X-ray emission could originate from a larger structure surrounding the inner torus. The neutral Fe Kα line emission is clumpy, with an equivalent width varying spatially from 0.5 to 3 keV, on a scale of tens of parsecs. This spatial variation in the Fe Kα line emission may be due to the ionization state and orientation effects of the reprocessing material, with respect to the incoming AGN photons. An Fe XXV He-α complex is also detected, consistent with photoionization from the AGN (Marinucci et al. 2017). A similar flattened distribution (∼160 pc across; Marinucci et al. 2013; Arevalo et al. 2014) is found in the next closest AGN imaged with Chandra, the Circinus galaxy at a distance of ~4.2 Mpc.

Spatial structures in the neutral Fe Kα line emission that can be related to nuclear disks or rings have been reported in two more CT AGN: NGC 5643 (Distance ~ 17 Mpc) and NGC 2110 (Distance ~ 33 Mpc). In NGC 5643 (Fabbiano et al. 2018c), a high-resolution Chandra image of the in the nuclear region revealed a clumpy neutral Fe Kα feature extending for ∼65 pc north to south (N-S), with no counterpart in the 3.0 - 6.0 keV continuum (Figure 6-16, left). This Fe Kα feature is spatially consistent with the N-S elongation of the rotating molecular disk of 26 pc diameter found in the CO(2-1) transition with ALMA in this galaxy (Alonso- Herrero et al. 2018), although slightly more extended. Both the CO and the Fe Kα emission could originate from a nuclear obscuring disk or torus. In NGC 2110 (Kawamuro et al. 2020), a comparison of Chandra and ALMA data shows corresponding extended Fe Kα and CO line emission that could be part of a system of nuclear rings. In both the nuclear disk of NGC 5643 (Fabbiano et al. 2018c) and in the nuclear features of NGC2110 (Kawamuro et al. 2020), the Fe Kα equivalent widths are found to be in the range of ~1-2 keV, consistent with fluorescence from nuclear X-ray irradiation interacting with dense obscuring clouds.

High resolution Chandra imaging of the Fe Kα emission of ESO 428-G014 (Distance ~ 23 Mpc) provides another example of clumpy emission in the inner ~30 pc, which could -at least in part - be associated with the obscuring torus (Fabbiano et al. 2019a). Although there is no definitive X-ray signature of a flattened Fe Kα line emission nuclear structure, ALMA observations of this galaxy (Feruglio et al. 2020) suggest a warped disk or bar the inner 100 pc region and fast gas streams which may trace an inflow toward the AGN. As in the case of NGC 5643 (Alonso-Herrero et al. 2018) the column density derived from the molecular cloud observations is ~few x $10^{23}$ $cm^{-2}$. While smaller than the nuclear obscuration of a CT AGN (~$10^{24}$ $cm^{-2}$), these results show that the circumnuclear dense molecular clouds can substantially contribute to the AGN obscuration.

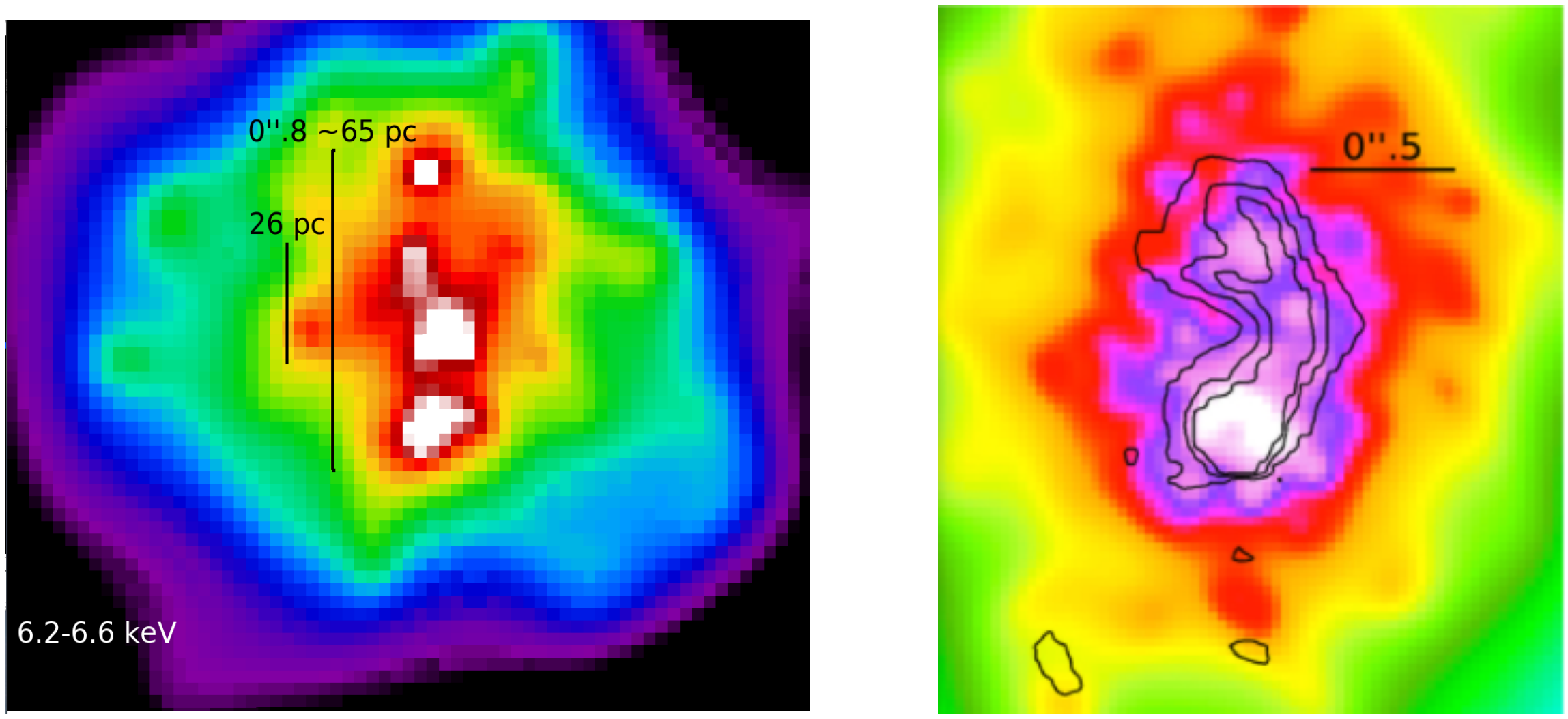

Figure 6-16: *Left*, Adaptive smoothing of the central north-south Fe Kα structure discovered with Chandra in NGC 5643 (Fabbiano et al. 2018c; white is highest intensity); this structure extends in the same direction as the CO (2-1) line emission and the central rotating CO (2-1) 26 pc disk discovered with *ALMA* (Alonso-Herrero et al. 2018). *Right*, Central high surface brightness region of NGC 2110 in the 0.1–1 keV band, obtained with 1/16 pixel binning and adaptive smoothing (Fabbiano et al. 2019b). Superimposed are the HST Hα+[N II] contours derived from Figure 2 of Rosario et al. (2019), which reside in the CO (2-1) cavity.

Comparisons of Chandra and ALMA images have also provided direct evidence of the effect of dense X-ray photon fields in warming the cold molecular clouds around the AGN, resulting in a localized lack of CO 2-1 emission where the extended X-ray emission is most intense. CO 2-1 ‘cavities’ or ‘lacunae’ have been found in the X-ray bright nuclei of NGC 1068 (Garcia Burillo et al. 2010) and Circinus (Kawamuro et al. 2019). X-rays filling these CO 2-1 cavities have been reported in NGC 2110 (Rosario et al 2019; Fabbiano et al 2019b; Kawamuro et al. 2020; see Figure 6-16, right) and in ESO 428-G014 (Feruglio et al. 2020), where there is also evidence of warmer clouds in the cavity emitting in $H_2$ and optical lines. The X-ray radiation field may alter the chemistry of these regions. Alternatively, fast X-ray producing shocks may warm up the CO molecules resulting in brighter emission in higher-J lines (see e.g., Feruglio et al 2020 and Kawamuro et al. 2020). The effect of X-ray excitation of $H_2$ lines and suppression of CO emission has also been discussed in the context of NGC 4151 (Storchi-Bergman et al. 2009; Wang et al. 2011a).

## 6.8 MAPPING THE PAST HISTORY OF AGNS

The time variability of AGNs gives important information about the physical properties of the black hole accretion disks and on the accretion history of the supermassive black holes at their centers (e.g., Czerny et al. 2006). Viscous timescales in alpha accretion disks are thousands of years. However, direct time variability studies of AGNs are limited to maximum scales of decades by the availability of appropriate observing facilities. While many interesting observations of variability have been made on decade timescales (Lawrence 2018), the longer lifetimes of AGNs are expected to be in the $10^7$ yr range (Martini 2004). X-ray mapping of the extra-nuclear emission of AGNs instead allows us to explore AGN variability on longer time scales, from the ~100 yr variability timescales of the nucleus of the Milky Way (Sgr A*), which hosts a black hole of $4.5 \times 10^6$ $M_{sol}$ (Ghez et al. 2008) to $10^{4-5}$ yr in nearby AGNs.

In the SgrA* region diffuse emission in the 2-10 keV band and in the 6.4 Fe Kα line was discovered with the Japanese satellite ASCA (Koyama et al. 1996). This emission is akin to the extended hard continuum and Fe Kα emission found in several nearby AGN with Chandra (see Section 6.5) and is likely to be the result of the hard X-ray photons from the AGN interacting with the surrounding dense molecular clouds. In the Sgr A* region, an Fe Kα emitting cloud was discovered with ASCA at ~90 pc from Sgr A*. The luminosity of this cloud implies an incident nuclear luminosity $>10^4$ times greater than the present luminosity of the Sgr A* X-ray source. Given the light travel time, this luminous cloud indicates a major nuclear flare occurring ~300 yr ago.

More recent comparisons of Chandra images of the Sgr A* region show variability in both spatial and spectral features in the region ~100 pc from the black hole, consistent with the X-ray echo of a nuclear flare. Churazov et al. (2017a, 2017b) present a time-evolution model of the X-ray morphology, assuming a factor of 10 Sgr A* outburst (from $L_X = 10^{40}$ erg $s^{-1}$) lasting for 50 or 5 yr. The illuminated regions would be visible for the first 200 yr, before becoming diffuse and disappearing, with a time evolution governed by the distribution of the clouds and their densities. Fabbiano et al. (2019a) suggest that a similar effect could be responsible for the two nuclear Fe

Kα knots observed in the CT AGN ESO 428-G014. The ~30 pc projected separation of these Fe Kα knots constrains the light travel time (and therefore the occurrence of the flare) to be >~90 yr.

The hot bubble of NGC 4151, discovered with Chandra, is a ~3 kpc radius region of diffuse X-ray emission (hot gas, the extended blue region in Figure 6-16), that appears to fill the HI cavity in this galaxy (cold gas, the red region in Figure 6-16) surrounding the nucleus (Wang et al. 2010). Convincing evidence points to a physical link of this radiation to the AGN activity. If the ISM in the bubble is photoionized, its physical parameters imply an AGN Eddington limit outburst < 1.5 x $10^4$ years ago. If, instead, it is a hot optically thin thermally emitting region, perhaps shock ionized by nuclear winds, then continuous heating for ~$10^5$ years can be inferred (Figure 6-17). Future X-ray observations of this region, with a telescope with angular resolution comparable to Chandra, but a significantly larger collecting area and spectral resolution will be needed to discriminate spectrally between the photoionization and thermal hypotheses.

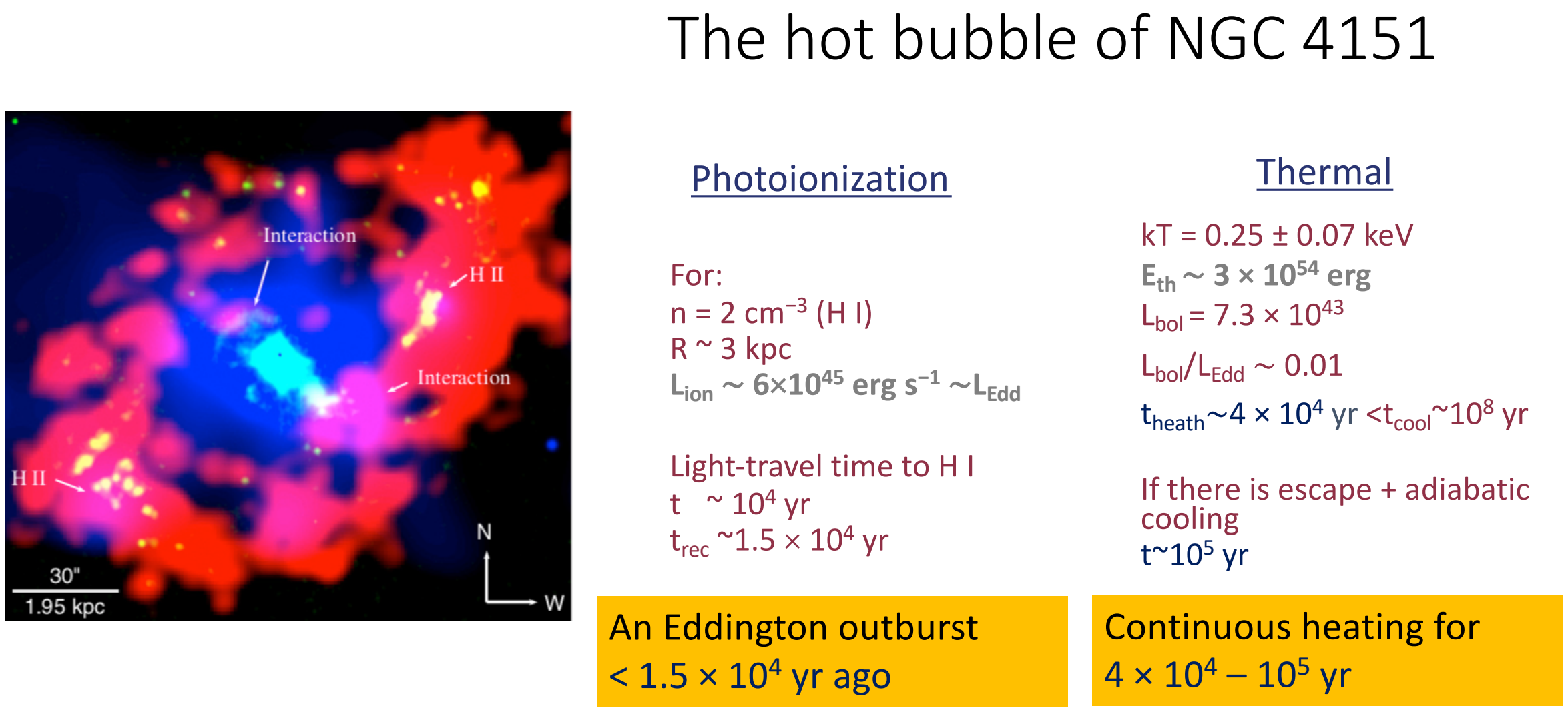

Figure 6-17: *Left*, The hot circumnuclear 2 kpc radius X-ray bubble of NGC 4151 (blue), within the ring of HI emission, in red. Star-forming HII regions are shown in yellow. *Right*, Summary of constraints on the time-evolution of the hot bubble of NGC 4151 (adapted from Wang et al. 2010).

The kiloparsec-extended bicones frequently associated with AGNs (Section 6.4) suggest a prolonged nuclear activity and interaction with the ISM. For example, the ~5 kpc bicone of ESO 428-G014, based on light travel time, implies at least quasi-continuous activity for ~$10^4$ yr (Fabbiano et al. 2019a). Activity times of ~$10^4$-$10^5$ yr have also been reported, based on the large-scale ionization cones, for NGC 5252 (Dadina et al. 2010) and IC 2497 (Fabbiano & Elvis 2019). These outburst durations may suggest a period of high accretion rates, at least Eddington, for the AGN. See for example the comparison with the models of Czerny et al. (2009) in the case of ESO 428-G014, given the black hole mass estimate of a few $10^7$ solar masses (Fabbiano et al. 2019a).

In some galaxies these continuous nuclear outbursts may now have ended. A good example is provided by IC 2497. This is a quiescent galaxy (not a strong AGN), but has an extended [OIII] emission region, typical of ionization cones ~20 kiloparsecs to South of the nucleus; this region is known as "Hanny's Voorwerp"[4]. Given the intensity of this [OIII] emission and its expected lifetime, this discovery pointed to a past quasar outburst (Lintott et al. 2009). While subsequent X-ray observations found that IC 2497 is a CT AGN, they also confirmed that its current bolometric luminosity is significantly lower than that implied by the [OIII] feature (Sartori et al. 2018). A similar past brightening of the nuclear black hole may be responsible for the extended bicones of NGC 5252 (Dadina et al. 2010).

In IC 2497, the Chandra data showed a soft X-ray counterpart of the [OIII] ionization cone (Fabbiano & Elvis 2019). Both the [OIII] and the X-ray extent are consistent with a light travel time of $10^5$ yr. However, the [OIII] emission is fading fast, given the short [OIII] recombination time (~$10^4$ yr; Lintott et al. 2009), while the X-ray emission will persist for ~$10^7$ yr (Fabbiano & Elvis 2019). This indicates that while the [OIII] emission is a diagnostic of a time-limited outburst, the X-ray emission may be a better indicator of the total AGN activity. X-ray voorwerps (or voorwerpen in Dutch) should be more common than [OIII] voorwerps.

## 6.9 AGN FEEDBACK ON THE HOST GALAXY ISM

From the point of view of AGN feedback on the galaxy, X-ray imaging spectroscopy with Chandra puts some constraints on all three forms of feedback: jets, winds, and radiation.

For relativistic *jets*, the fraction of jet power deposited in the hot ISM for MRK 573 can be estimated from the X-ray emission at the ends of the jets where shock heating seems to be occurring. The fraction is large, ~28%, though the power of the jet, $P_{jet}$ ~$10^{42}$ erg $s^{-1}$ is just ~0.2% of the bolometric luminosity of MRK 573 (Paggi et al., 2012). In radio quiet AGN this is not an unusually low $P_{jet}/L_{bol}$ fraction. To make this estimate Paggi et al. (2012) use the thermal energy in the hot collisional ISM components, $E_{KE}$~$10^{54}$ erg, and the sound speed crossing time of the radio knots, $t_{th}$~$10^5$ yr, to derive a kinetic luminosity $L_{KE}$~$10^{41}$ erg $s^{-1}$. As the AGN bolometric luminosity of MRK 573 is $L_{bol}$ ~ $10^{44}$ erg $s^{-1}$, $L_{KE}/L_{bol}$ ~0.05%. For NGC 4151 the same arguments for the X-ray hotspots at the ends of the radio jets, particularly the strong Ne IX/O VII region NE of the nucleus, ~100 pc from the nucleus leads to the conclusion that 0.1% of the jet power is being deposited in the ISM (Wang et al. 2011b). Larger samples are needed to find a reliable mean $L_{KE}/L_{bol}$. Nardini, Kim, and Pellegrini (2022, Chapter 4, this volume), cover the case of radio loud AGN.

For winds or *outflows*, the soft X-ray observations suggest that a small fraction of the AGN power directly impacts the host galaxy ISM in the form of the kinetic energy of the outflows.

[4] The [OIII] feature was discovered by Hanny van Arkel in the vicinity of IC 2497. Ms. Van Arkel was one of 100,000 participants in the Galaxy Zoo project, which aimed at classifying morphologically ~900,000 objects from the Sloan Digital Sky Survey; 'voorwerp' is Dutch for 'object' (Lintott et al. 2009).

Wang et al., (2011c) estimate the kinetic luminosity of the kT ~1 keV X-ray outflow in NGC 4151. Good "hollow cone" kinematic models for the outflow exist for this AGN (Das et al., 2005, Storchi-Bergmann et al., 2010). The optical/UV-based outflow rate is ~0.16 $M_{sol}$/yr (Crenshaw & Kraemer 2007), while Wang et al. (2011c) find that the X-ray gas gives a ~12 times higher rate, ~2.1 $M_{sol}$/yr, assuming the same outflow velocity (650 km $s^{-1}$). This X-ray rate is comparable to the near-IR (ZJHK bands) rate of ~1.2 $M_{sol}$/yr based on IFU data (Storchi-Bergmann et al. 2010). The X-ray kinetic luminosity is then ~4 x $10^{41}$ erg $s^{-1}$, some 0.4% of the AGN bolometric luminosity of NGC 4151 ($L_{bol}$ ~ $10^{44}$ erg $s^{-1}$, Kaspi et al. 2005). Adding in the near-IR component, this could reach ~0.7%.

In MRK 78, with $L_{bol}$ ~ (1 - 4) x $10^{44}$ erg $s^{-1}$, the deceleration of the NLR outflow leads to X-ray emission of ~(2 - 6) x $10^{41}$ erg $s^{-1}$, giving $L_{KE}/L_{bol}$ ~(0.05 - 0.6)% (Fornasini et al. 2022). MRK 78 has the X-ray emission offset downstream from the deceleration location by ~1 kpc. This is expected for a shock in a normal ISM density of ~1 $cm^{-3}$ if the shock temperature is too high for there to be strong line emission, as seen in Herbig-Haro outflows (Heathcote et al. 1998). Bremsstrahlung alone is an inefficient radiator, so the hot gas travels a long way before it cools enough to emit strongly via emission lines. This effect complicates the measurement of $L_{KE}$ (Fornasini et al. 2022).

In some model outflows, carrying 0.5% of $L_{bol}$ is sufficient to suppress star formation in the host by disrupting the molecular clouds where that formation takes place (Hopkins & Elvis 2010). The measured feedback power lies interestingly close to this critical value below which it has negligible effect. More estimates of the total (optical-IR + X-ray) $L_{KE}/L_{bol}$ would be valuable.

For *radiation*, the hard extended X-ray emission in ESO 428-G014 and other AGN points to radiation driven heating of molecular clouds that may lead to their destruction on both small scales (~100 pc, Section 6.7) and on larger scales (~1 kpc, Section 6.5). The presence of CO 2-1 'cavities' where hard X-rays are seen could be an indication of this process (Section 6.7). The effects of X-rays on molecular clouds are complex to model (Yan & Dalgarno, 1997).

It is important to realize that only nearby and typically moderate luminosity CT AGNs are susceptible to being imaged with Chandra. Most have accretion rates a few percent of the Eddington limit. Much higher and lower Eddington ratios may well have different feedback effects (see Nardini, Kim, and Pellegrini, Chapter 4, this volume). Absorption line studies of low obscuration AGN by Fiore et al. (2017) find that at $L_{bol}$ ~ $10^{45}$-$10^{46}$ erg $s^{-1}$, the wind kinetic energy is comparable to the values found from imaging, i.e., ~1% of $L_{bol}$, albeit with an order of magnitude spread. Larger bicones are seen in high luminosity AGNs (e.g., Trindade Falcão et al. 2021) where Fiore et al. (2017) suggest wind kinetic energies may carry a larger fraction of $L_{bol}$. Complementary studies in X-rays will have to wait for sub-arcsecond imaging X-ray telescopes with much larger mirrors than Chandra.

## 6.10 SUMMARY - REVISED VIEW OF AGNS AND THEIR INTERACTION WITH THE HOST GALAXY

The combined use of Chandra's ultimate sub-arcsecond angular resolution, enabled by sub-pixel binning of ACIS images, together with the ACIS spectral capabilities has effectively created an X-ray instrument of comparable power to the current generation of optical-IR IFUs (VLT, Gemini) and of millimeter (ALMA) and radio telescopes (VLA). This approach has been used to investigate in detail the X-ray properties of about two dozen nearby radio-quiet obscured AGNs (see Section 6.4, Table 6-1 and Figure 6-3.) These Chandra highest-resolution, spectrally resolved, images show multiple X-ray components and striking correlations with ground-based multi-wavelength data. They are leading to a better understanding of both the properties of AGNs and the complex interaction of the AGN with its host galaxy, commonly known as “feedback”.

In summary, recent Chandra observations, in combination with similar angular resolution multi-wavelength data, have shown that:

- Galaxy scale (kpc) soft (<2 keV) X-ray emission is prevalent in CT AGNs (Section 6.4.3):
    - The extended soft X-ray emission has a collimated morphology;
    - The X-ray structures align with the optical emission line ionization cones;
    - X-ray bright spots are common at the ends of the small radio jets. The X-ray temperatures then give the speed of the shocks.

- The ‘Seyfert’ emission regions are extended and physically complex (Section 6.4.3, 6.4.5):
    - Comparison with Hubble optical line excitation maps have shown that extended ‘Seyfert’ emission regions are mostly correlated with the intense X-ray emission features, supporting an AGN photoionized picture.
    - Spectral analysis of the ACIS data of these regions requires complex multi-component models spanning a range of ISM photoionization plus thermal (shock?) emission.

- X-ray emission line imaging (Section 6.4.4):
    - Spatial variations in line ratios are common, pointing to varying physical conditions across the structures.
    - Localized excesses of Ne IX-X emission may be associated with regions shock ionized by the impact of AGN jets.
    - Localized differences in these narrow line images provide a direct diagnostic of different physical properties.

- LINER emission regions are commonly found even in galaxies classified as pure AGNs from their integrated nuclear spectrum (Section 6.4.6):
    - Extended LINER regions found in HST BPT imaging often surround the soft extended X-ray ‘Seyfert’ emission.
    - This morphology suggests a shock origin for the long-ambiguous LINERs, as the X-ray wind expands into the ISM.

- Galaxy scale hard (~3-6 keV) continuum X-ray emission is also present in at least some CT AGNs together with 6.4 keV neutral Fe Kα line emission (Section 6.5):
    - In some cases, > 30% of the observed hard emission is extended, suggesting that it is the result of the interaction of AGN photons with molecular cloud complexes in the host. Instead, the extended >2 keV emission must be taken into account

when modeling broad-band AGN X-ray spectra or the torus parameters will be biased.
  - In the well-studied case of ESO428-G014, the luminosities of the soft (0.3-3 keV) and hard (3 -7 keV) emission, from a region of 1'.5-8" around the nucleus are similar, ~$10^{40}$ erg $s^{-1}$. Since there is likely to be, so far unresolved, extended emission even closer to the nucleus, these estimates should be considered lower limits.
  - These components were previously assumed to be associated with the AGN itself and its parsec-scale torus.
  - AGNs interact with the molecular clouds near the nucleus of the host via radiation and outflows.
  - In the highly disturbed double AGN merger NGC 6240, both thermal and non-thermal >2 keV extended emission features suggest both shock ionization and reflection from the dense molecular clouds seen with ALMA (Section 6.5.1).

- <u>There is emission in the cross-cone direction, which should be blocked by the nuclear torus in the standard model</u> (Section 6.6):
  - A clumpy "porous torus" is one explanation; in this case the torus transmits ~1% - ~20% of the nuclear radiation.
  - At the softer energies, the emission may be due to a gravitationally trapped hot halo (see Bogdan and Vogelsberger, Chapter 4). However, interaction of the relativistic jets with the dense ISM, as shown by recent simulations, is also a possibility, supported by some recent evidence in high resolution emission line imaging.
  - In a few low luminosity cases X-ray binaries could account for the cross-cone extended X-ray emission.

- <u>AGN – molecular cloud interactions are seen in the central ~100-200 pc (Section 6.7):</u>
  - High resolution Chandra imaging in the 6.4 keV neutral Fe Kα line has uncovered extended features in the central ~100-200 pc that are related to extended outer structures of the obscuring torus and molecular clouds structure seen with ALMA (e.g., NGC4945, Marinucci et al 2012; NGC 5643 Fabbiano et al., 2018c).
  - Comparison of Chandra and molecular line data has shown directly how X-ray irradiation depletes the CO 2-1 emission in the central ~100 pc X-ray intense regions (ESO 428-G014, Feruglio et al. 2020; NGC 2110, Fabbiano et al., 2019b).

- <u>The past ~$10^4$ - $10^5$ yr of AGN activity can be inferred</u>, based on light travel time, energetics, and lifetime of the emission considerations (Section 6.8).

- <u>AGN feedback on the galaxy by all three mechanisms of jets, outflows and radiation is seen (Section 6.9):</u>
  - Estimates of the feedback power are a small fraction of the AGN bolometric luminosity (~0.5 % in NGC 4151).

- The feedback power lies interestingly close to a critical value below which it has negligible effect.
- The effect of the nuclear irradiation of molecular clouds near the nucleus, as shown by the hard X-ray extent, has not yet been carefully studied.

In conclusion, the emerging view is that the 'standard AGN model' of an accreting massive black hole obscured by a parsec-scale torus (e.g., Urry and Padovani 1995) is an oversimplified picture of the AGN phenomenon, which instead involves much larger, galaxy-scale, structures in the host (Figure 6 – 18).

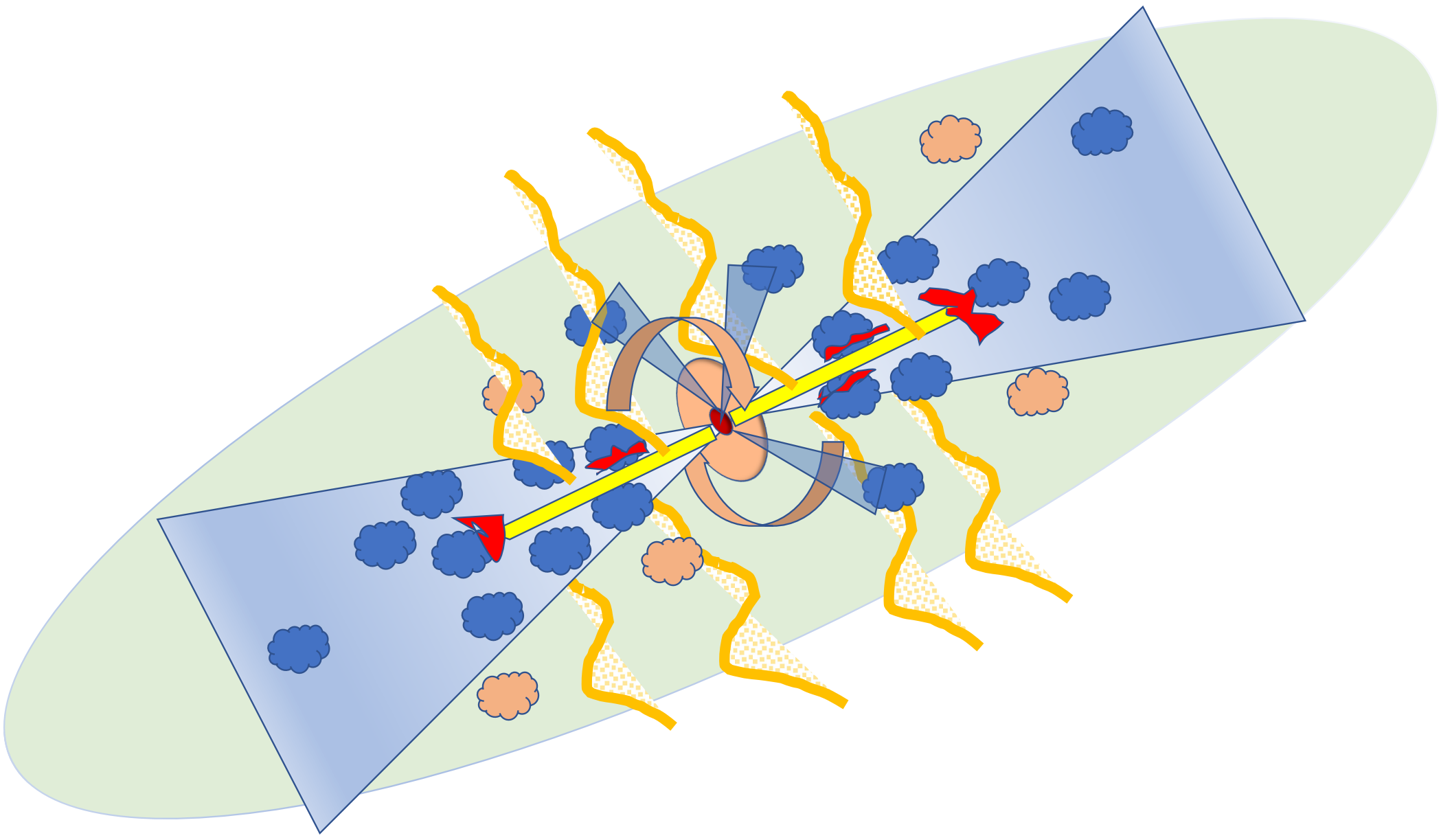

Figure 6-18: A cartoon representation of the main features discussed in this Chapter. Note: none of the features in this cartoon are on scale. The pale green ellipse represents the tilted galaxy disk; the pink clouds, central "spiral" streamers and disk represent molecular clouds, nuclear infalling molecular material and the circumnuclear molecular disk structure. The nuclear torus is represented by the red ellipse at the center of the large molecular disk. Irradiated molecular clouds emitting X-ray hard continuum and Fe K$\alpha$ are in blue. The pale blue triangles represent regions where AGN photons escape into the surrounding ISM, causing the ionization cones and partly the cross-cone ionized regions. Radio jets are in yellow and jet-ISM shocks in red. The orange tendrils in the cross-cone direction indicate the hot lateral outflows that may be partly responsible for the X-ray emission in these areas.

These results discussed in this Chapter demonstrate that X-rays data, of commensurate resolution to multi-wavelength images, are needed to develop a complete picture of AGN/host interaction. Along with radio continuum, mm and sub-mm molecular line emission, and optical/near-IR emission lines, these multi-wavelength data provide direct evidence of the morphology of the circum-nuclear region in 10-100pc scale, of gas inflows and outflows, and of the interaction of the nuclear radiation, winds and jets with the host galaxy ISM. Combined with the inflows ("feeding") seen in $H_2$, CO emission lines in the millimeter and optical/NIR we can start to construct a physics-based model of feedback.

The finding of hot and highly photoionized gas on 10s parsecs to several kiloparsec scales demonstrates that all three feedback mechanisms are at work: radiation affects the inner molecular clouds of the host on a ~1 kpc scale; shocks of relativistic jets with the host ISM a few kpc from the central AGN; and photoionization of the ISM by the nucleus on scales from pc to multiple kpc.

These results give us a tantalizing glimpse of what a future high resolution powerful X-ray telescope may reveal about the detailed physical interactions of supermassive black holes and galaxies. Observations of extended AGN X-ray emission are presently limited by S/N, which limits both surface brightness sensitivity and the number of emission lines that can be imaged; and by the limited energy resolution, $E/\Delta E \sim 10$, of silicon-based detectors. There are highly promising technologies that could deliver ~100X larger area mirrors without degrading the angular resolution, and for ~10 - 100X better spectral resolution detectors to go at the focus of these mirrors (Vikhlinin, Özel, and Gaskin 2019, Bandler et al. 2019, Zhang 2019).